\documentclass[conference]{IEEEtran}

\usepackage{graphicx}
\usepackage[
  ocgcolorlinks=true,
  linkcolor=blue,
  pdfauthor={David E. Robillard},
  pdftitle={Adaptive Software Radio Steganography},
  pdfsubject={Radio Steganography},
  pdfkeywords={Software, Radio, Steganography, Constellation},
]{hyperref}
\usepackage{url}
\usepackage{subfigure}
\usepackage{flushend}

\begin{document}

\title{Adaptive Software Radio Steganography}

\author{
  \IEEEauthorblockN{David~E.~Robillard}
  \IEEEauthorblockA{
    School of Computer Science, Carleton University\\
    1125 Colonel By Dr, Ottawa ON \, K1S 5B6, Canada\\
    \href{mailto:drobilla@scs.carleton.ca}{drobilla@scs.carleton.ca}
  }
}

\maketitle

\begin{abstract}
  This paper presents an adaptable steganography (information hiding) method for digital radio communication.
  Many radio steganography methods exist,
  but most are defined at higher levels of the protocol stack and are thus protocol dependent.
  In contrast, this method is defined at the physical layer,
  which makes it widely applicable regardless of the protocols used at higher layers.
  This approach is also adaptive; the covertness of the hidden channel is simple to control via a single continuous parameter either manually or automatically.
  Several variations are introduced,
  each with performance evaluated by simulation.
  Results show this to be a feasible method with a reasonable trade-off between performance and covertness.
\end{abstract}

\section{Introduction}

Steganography is the art and science of hiding one message in another such that an observer does not suspect the presence of hidden information.
The hidden message is called the \emph{secret}, and the message in which it is hidden is called the \emph{cover}.
A receiver that is not aware of the steganographic channel is called a \emph{legacy} receiver.

Steganography methods exist for most media, such as images or audio, including analogue and digital radio.
This paper addresses steganography in digital radio transmissions,
where information is transmitted as a sequence of \emph{symbols}.
A symbol represents $n$ bits and is encoded as a particular phase, amplitude, or frequency of the radio wave for a chunk of time.

The basic goal of digital steganography is to convey additional bits of information along with those of the cover.
This can be done at the symbol level: where the legacy signal has $2^n$ symbols each with $n$ bits,
secretly there are actually $2^{n+k}$ symbols each with $n+k$ bits.
The secret symbols are defined in such a way that a legacy receiver can correctly decode the $n$ bits of the cover.

The number of bits per symbol and precise way each symbol is mapped to a signal vary in real-world systems,
but all digital radio is based on the same fundamental principles.
A steganography method defined at the physical layer can thus be used in almost any digital radio system,
e.g.\ 802.11 WiFi networks, digital mobile phones, digital television broadcasts, and so on,
though some modulation schemes are more suited to this modification than others.

\section{Background}

\subsection{Previous Work}

Several steganography methods~\cite{Banoci2009}\cite{Mehta2008} exist for spread spectrum technologies
like Direct Sequence Spread Spectrum (DSSS) and Code Division Multiple Access (CDMA),
where each $n$ bit symbol is multiplied by a longer pseudo-random \emph{chip code}.
This has the effect of \emph{spreading} symbols over more bandwidth, which adds error tolerance.
Chip codes are chosen such that spread symbols are well separated in hamming distance.
The receiver resolves symbols to the closest known symbol in hamming distance,
thus tolerating bit errors.
Since the spread symbols are much longer than $n$ bits,
this process creates room to hide a secret channel at the cost of some error tolerance.
The spreading is designed to tolerate bit errors,
but as a result secret information can be conveyed by {\em deliberately} flipping bits.
As long as the number of bits flipped is low enough that the manipulated symbol is closer to the correct cover symbol than any other,
a legacy receiver will work correctly.
A trivial example is to simply use the least significant bit of each spread symbol as a secret channel.
Practical schemes are more sophisticated, but based on the same general idea:
if the receiver can tolerate $b$ bit errors per symbol,
then with sufficiently good signal quality, up to $b$ bits can be deliberately manipulated to convey secret messages.

One disadvantage of chip code based steganography is that it only works in protocols that use such codes.
Many common protocols do,
but the details vary dramatically which makes adapting to a new protocol non-trivial.
Another important disadvantage is that it is difficult to generate a suitable set of steganographic symbols.
Discovering a superior such set is a proposed avenue of future research~\cite{Mehta2008},
not something simple to adjust dynamically.
The distance between symbols in the state of the art is fixed,
and is inherently integral (being defined by hamming distance) and of limited range
(being limited by the number of tolerable bit errors).
The intricate relationship between symbol distance and number of possible symbols adds additional complexity.

Hence, chip-code based methods are not adaptable:
the subtlety can not be easily controlled via a continuous parameter.
The ability to do so would be useful for achieving the most subtle secret channel possible that is feasible with the current signal quality.

Another strategy is to take advantage of \emph{white space}: blocks of time that are unused in a protocol due to packet padding~\cite{Szczypiorski2011}.
This unused time can be used to transmit secret messages.
Some protocols have enough white space to transmit at relatively high data rates,
e.g.\ WiFi, where a secret channel with a rate of over 1 Mbit/s is possible.
However, this method alone is not very interesting from a covert communication perspective since the secret channel would be obvious to any observer looking for it.
Where a normal transmission should be zero there would be a signal,
making it obvious that white space is being used for transmission.
This approach is suitable for achieving high data rates without interfering with legacy receivers,
but not where security is more important than performance.
Like chip code schemes, it is also intimately tied to a particular protocol since white space exists at very specific times defined in higher layers.
This method is also not particularly radio specific;
similar methods can be used in a purely software domain anywhere padding exists.

In contrast, the approach presented here is defined at the physical layer, protocol independent, radio specific, and adaptable.
The performance vs. subtlety trade-off can be easily tuned at run-time according to the current signal conditions,
either manually or automatically as part of a cognitive radio steganography system.
This is achieved by manipulating the \emph{constellation} which defines how symbols are mapped to a physical wave.

\subsection{Constellations}

\begin{figure}[t]
  \begin{center}
    \includegraphics[width=0.9\linewidth]{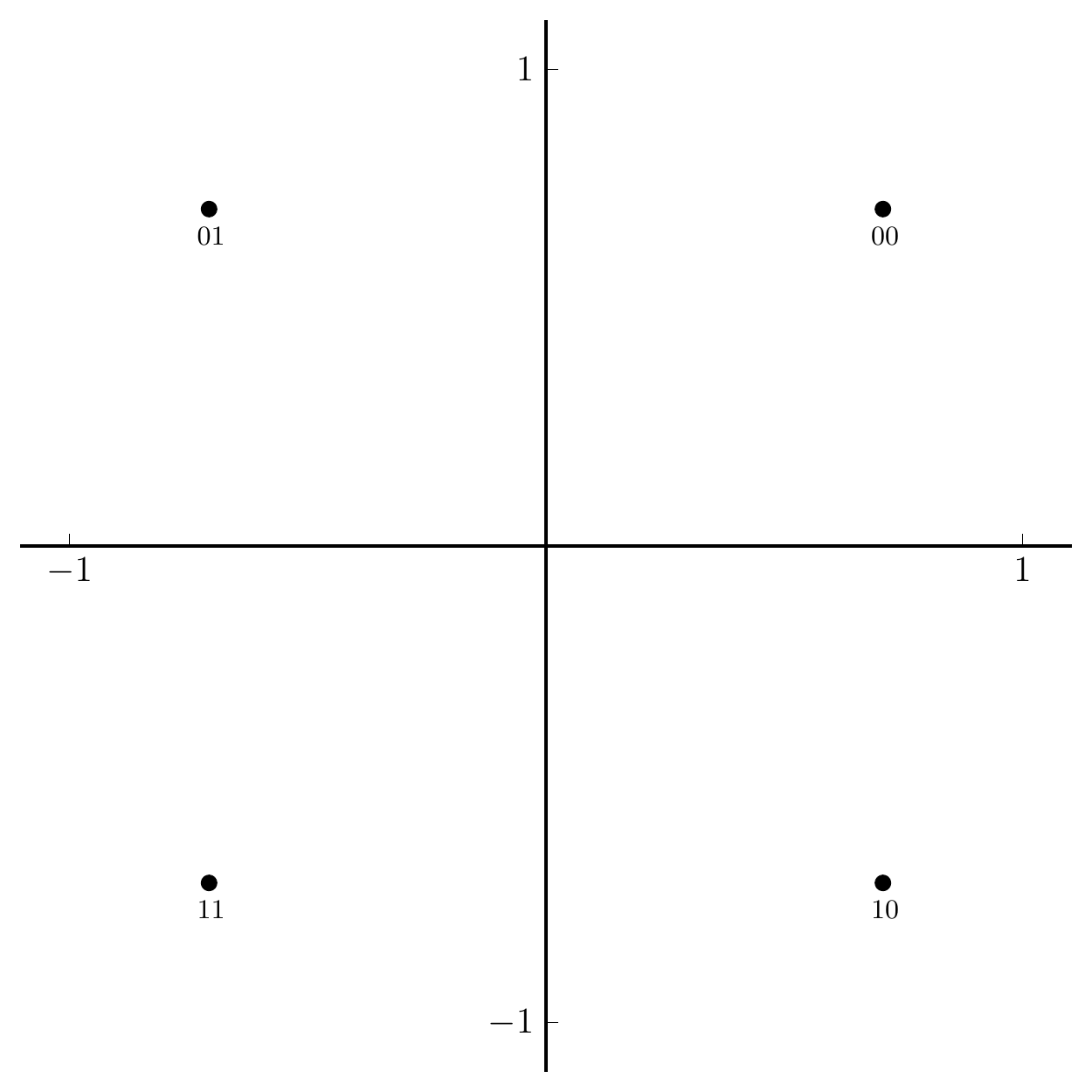}
  \end{center}
  \caption{Standard 4-QAM/4-PSK Constellation}
  \label{const_qam_n4_b1}
\end{figure}

\begin{figure}[t]
  \begin{center}
    \includegraphics[width=0.9\linewidth]{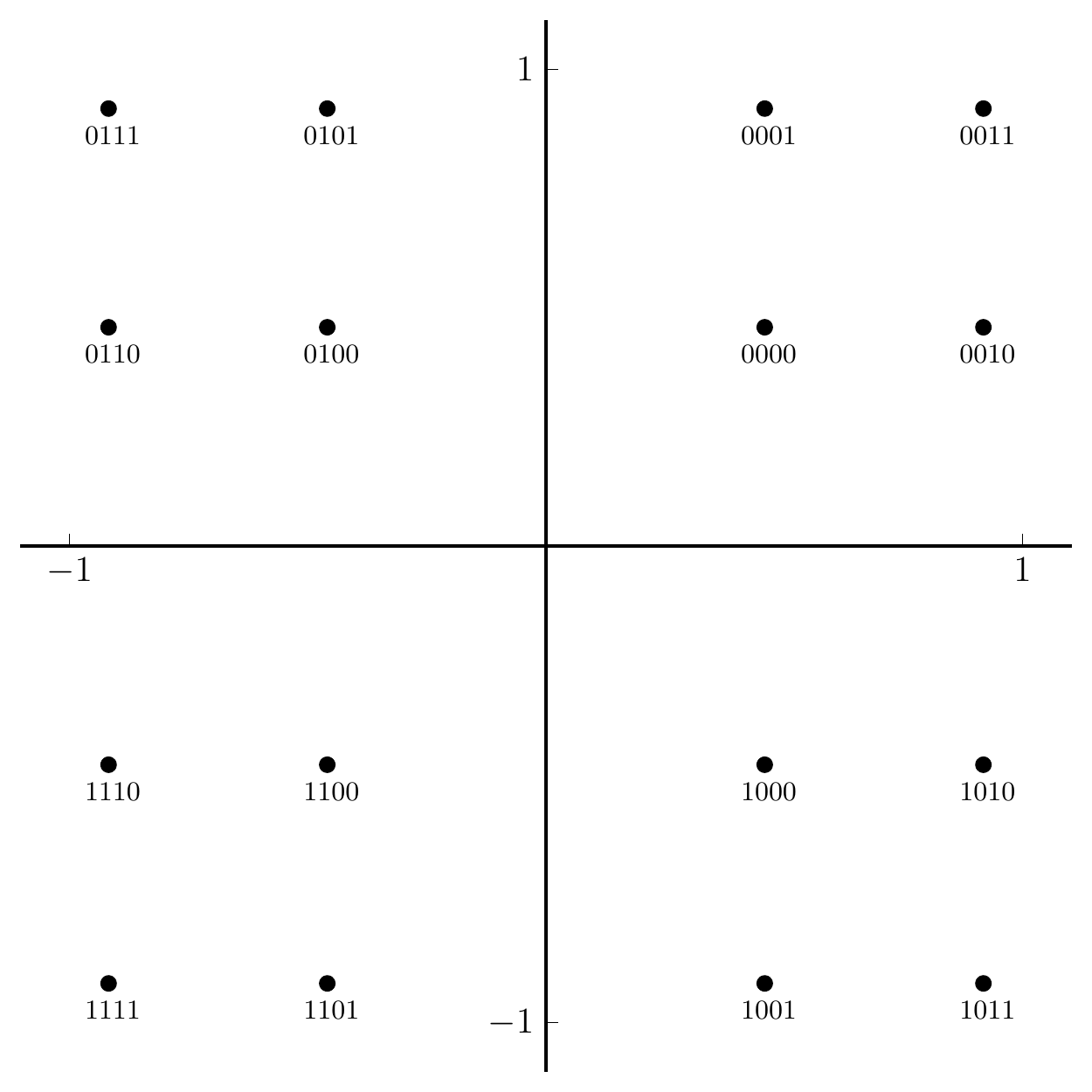}
  \end{center}
  \caption{Standard 16-QAM Constellation}
  \label{const_qam_n16_b1}
\end{figure}

\begin{figure}[t]
  \begin{center}
    \includegraphics[width=0.9\linewidth]{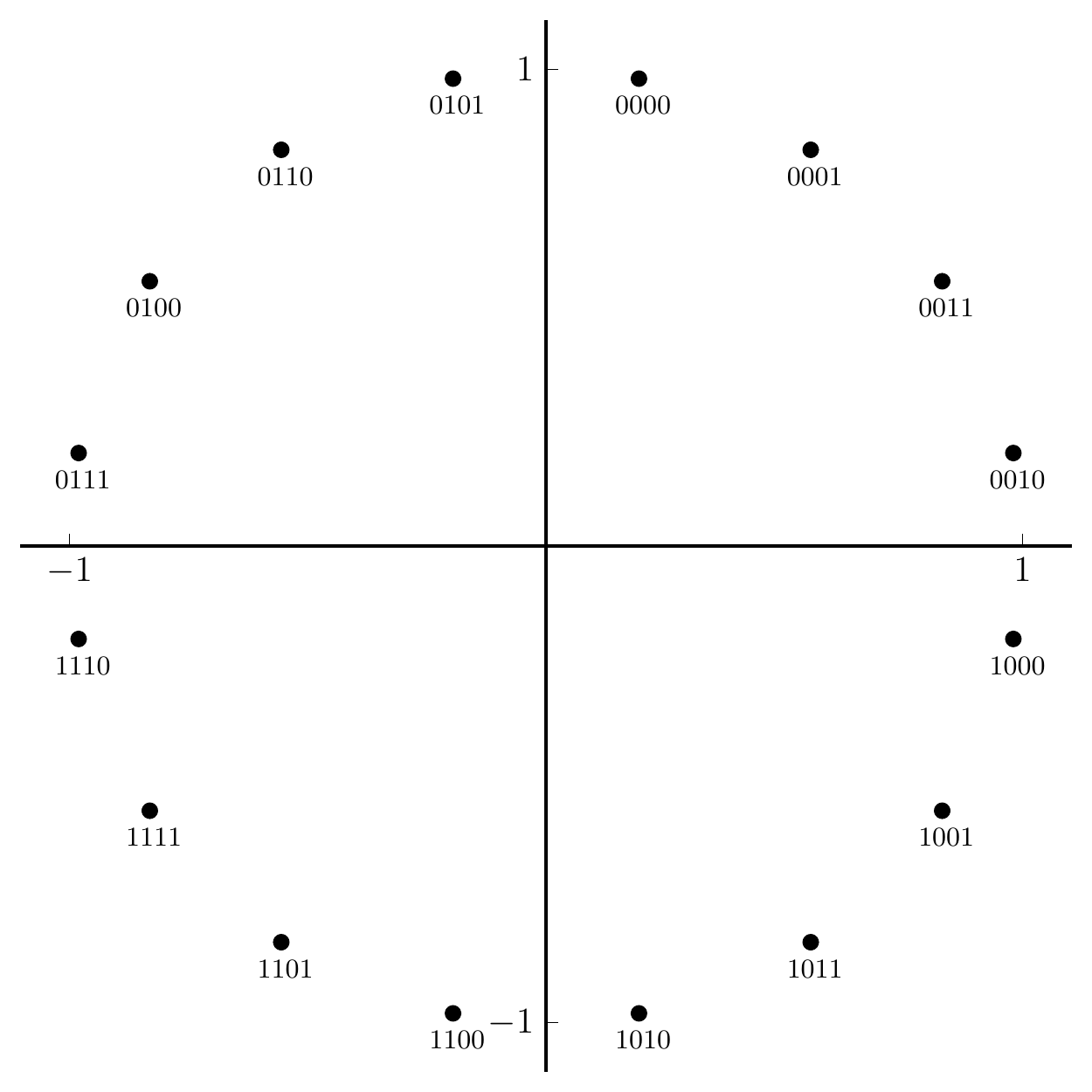}
  \end{center}
  \caption{Standard 16-PSK Constellation}
  \label{const_psk_n16_b1}
\end{figure}

Digital signals based on periodic waveforms, like radio waves, have 3 parameters suitable for modulation:
amplitude, frequency, and phase.
To encode digital information, a specific amplitude/frequency/phase is mapped to a symbol which represents $n$ bits.
Typically, frequency is used to divide the spectrum into channels,
while phase and amplitude are modulated to represent symbols
(frequency may also be modulated within a band).
The mapping of symbols to phase and amplitude can be illustrated with a constellation diagram,
where symbols are points,
the horizontal axis represents the \emph{in-phase} $I = \cos(x)$,
and the vertical axis represents the \emph{quadrature} $Q = \sin(x)$.
Thus the phase (angle from the positive $x$-axis) and amplitude (distance from the origin) of a point is apparent.
There are several standard ways to arrange $m$ symbols in a constellation,
where $m$ is usually a power of 2.
One arrangement for $m=4$, known as 4-QAM (for Quadrature Amplitude Modulation)\footnote{Specifically \emph{rectangular} QAM, there is also circular QAM but this paper only uses the rectangular variant and refers to it simply as ``QAM''.}
or 4-PSK (for Phase Shift Keying) is shown in Fig.~\ref{const_qam_n4_b1}.
PSK and QAM are identical for 4 symbols, but the difference can be seen for higher $m$.
A 16-QAM constellation is shown in Fig.~\ref{const_qam_n16_b1},
and a 16-PSK constellation is shown in Fig.~\ref{const_psk_n16_b1}.
The bits corresponding to each symbol are gray coded so that adjacent symbols differ by 1 bit,
which improves error tolerance.
As their names suggest, QAM modulates both the phase and amplitude of the signal,
while PSK modulates only the phase.
For each time slot, the receiver detects the phase and amplitude of the incoming signal,
and resolves to the closest symbol in the constellation.
Thus there is limited error tolerance built in at the physical layer,
though precisely how much is difficult to say as it depends on many factors.

\section{Approach}

The fact that the receiver correctly resolves symbol points that are slightly distorted can be exploited to hide a secret channel.
Around each symbol point, there is a region where any point resolves to that symbol.
If a cluster of $k$ secret points is arranged in this region,
they can be correctly resolved given sufficiently high signal quality.
Thus if cover symbols are $n$ bits,
there are a total of $2^{n + k}$ secret symbols, each with $n + k$ bits.
By choosing the most significant $n$ bits of each symbol to be the cover,
and the least significant $k$ bits to be the secret (or vice-versa),
a cover and secret can be transmitted simultaneously.
Since each of the $k$ symbols clustered around a legacy symbol is closer to that legacy symbol than any other,
a legacy receiver resolves only the cover bits, ignoring what seems like noise.
The same signal can be resolved by a steganographic receiver to the full $n + k$ bits per symbol.
Dividing these bits appropriately yields the original cover and secret bits.

Since symbols are defined as 2-dimensional coordinates in a continuous space,
it is simple to tune the secret constellation.
The \emph{blatancy} $\beta$ represents how spread out clusters are from the original symbol point.
It is normalised such that $\beta = 0$ produces all points exactly on cover points (making secret communication impossible),
and $\beta = 1$ produces the maximum reasonable spread,
usually equivalent to the standard constellation with the same number of symbols.

\subsection{Square Clusters}

\begin{figure}[t]
  \begin{center}
    \includegraphics[width=0.9\linewidth]{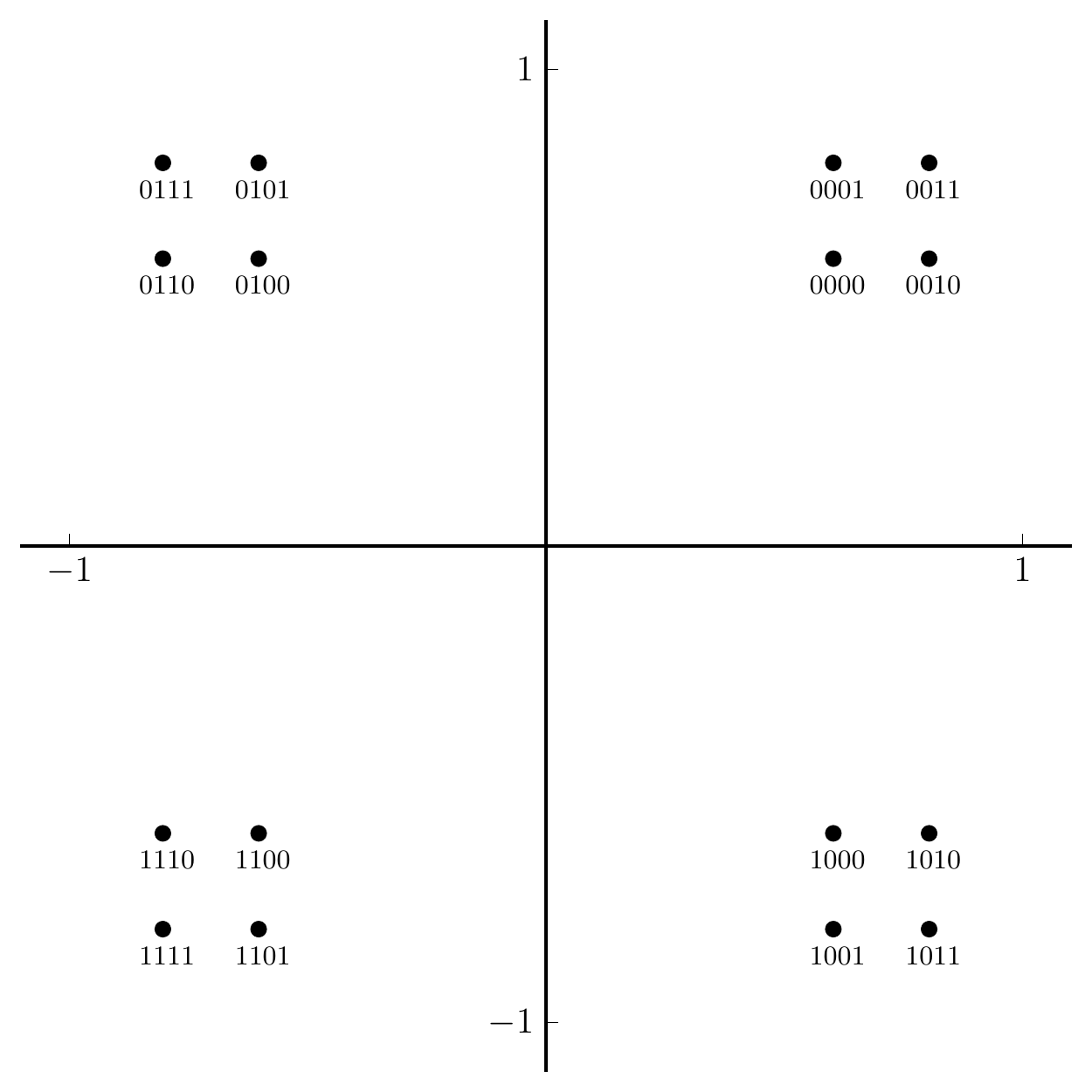}
  \end{center}
  \caption{Square 4x4-QAM Secret Constellation, $\beta=0.5$}
  \label{const_qam_n16_b05}
\end{figure}

A steganographic constellation based on QAM is shown in Fig.~\ref{const_qam_n16_b05}.
The cover constellation is 4-QAM as shown in Fig.~\ref{const_qam_n4_b1}.
There are clusters of 4 symbols arranged in a square around each legacy symbol,
where the most significant 2 bits of each is equivalent to the 2 bits of the cover symbol.
Hence any of these 4 symbols are resolved to the same 2 bits by a legacy receiver.
Since there are 4 secret symbols per cover symbol,
this arrangement is referred to as ``square 4x4-QAM''.
The blatancy controls the size of each square,
such that $\beta=1.0$ produces a standard 16-QAM constellation as shown in Fig.~\ref{const_qam_n16_b1}.
Lower values cluster points closer to the corresponding cover points,
which increases stealth but also reduces the reliability of the steganographic channel.

\subsection{Circular Clusters}

\begin{figure}[t]
  \begin{center}
    \includegraphics[width=0.9\linewidth]{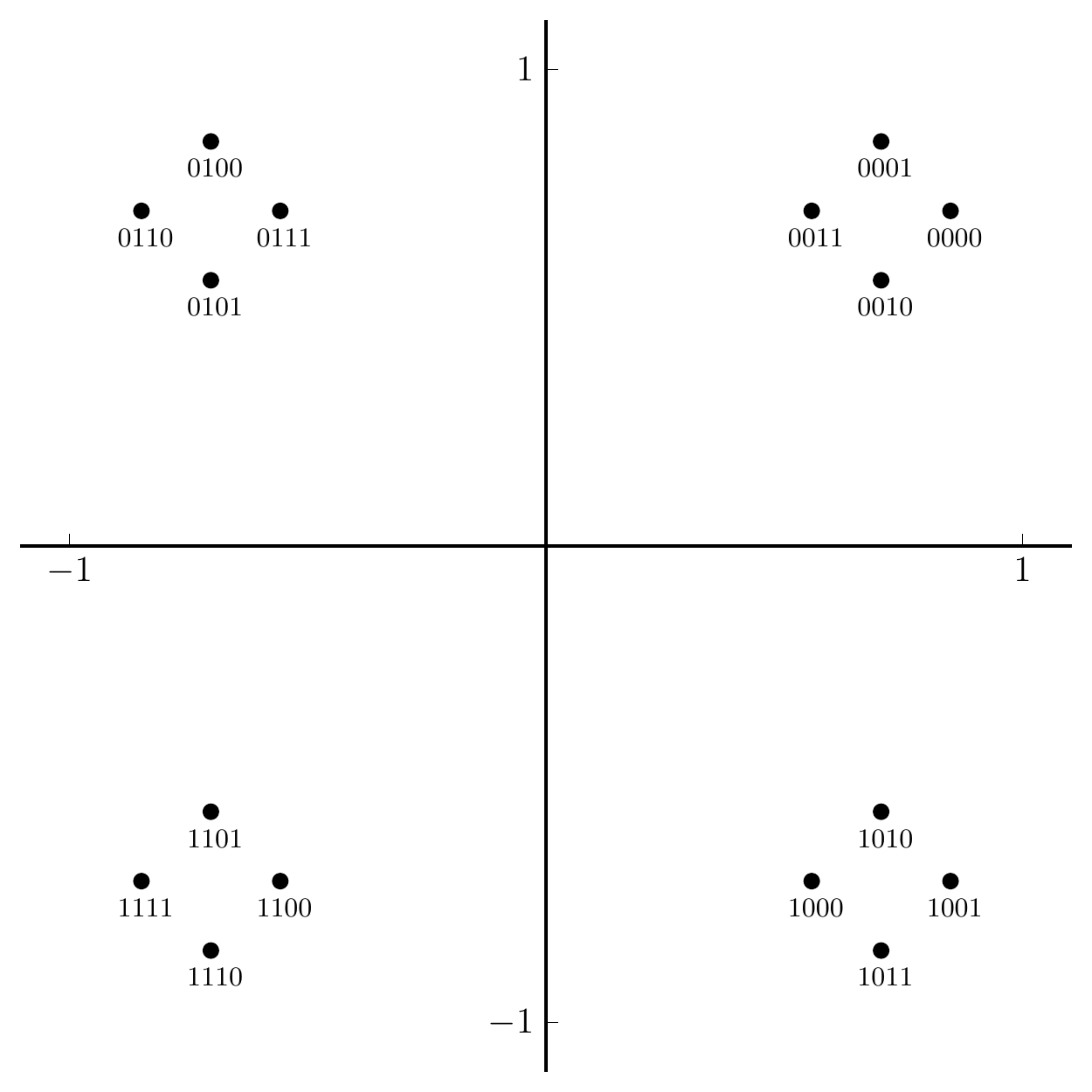}
  \end{center}
  \caption{Circular 4x4-QAM Secret Constellation, $\beta=0.5$}
  \label{const_cpsk_n16_b05}
\end{figure}

A similar scheme is to arrange secret symbols in a circle around the cover points,
as shown in Fig.~\ref{const_cpsk_n16_b05}.
This arrangement is referred to as ``circular 4x4-QAM''.
Like square 4x4-QAM, the blatancy controls the size of each cluster,
in this case the radius of the circle.
The maximum, $\beta = 1.0$, is the size where points become as close to points in other clusters
as they are to their own cover point.
This arrangement is very similar to square 4x4-QAM,
but circular clusters are more convenient for adding additional modulation to increase stealth.
Such an improvement is described in Section~\ref{stealth}.

\subsection{Phase Shift Keying}

\begin{figure}[t]
  \begin{center}
    \includegraphics[width=0.9\linewidth]{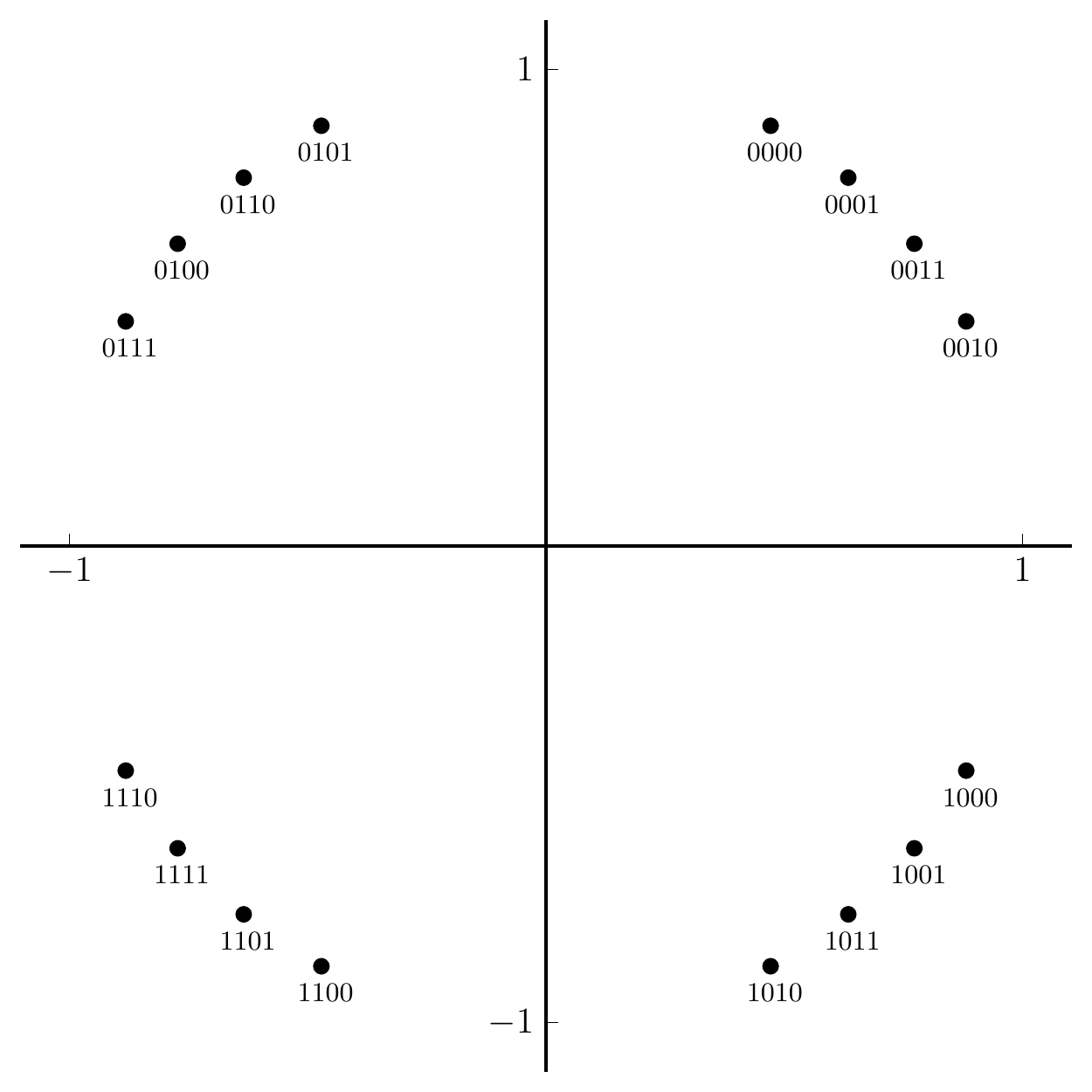}
  \end{center}
  \caption{Secret 4x4PSK Constellation, $\beta=0.5$}
  \label{const_ppsk_n16_b05}
\end{figure}

Some radio platforms use only PSK and thus are not designed to detect amplitude modulation.
In order to achieve steganography in this context,
a secret constellation based exclusively on phase modulation is required.
One such arrangement, referred to as ``4x4-PSK'', is shown in Fig.~\ref{const_ppsk_n16_b05}.
Here, $\beta$ controls the angular distance over which clusters are spread.
The maximum $\beta = 1.0$ is where all adjacent points are equidistant,
resulting in a constellation identical to standard 16-PSK as shown in Fig.~\ref{const_psk_n16_b1}.

\section{Performance}

In order to evaluate the performance of each constellation type,
a simulation was performed using GNU Radio~\cite{gnuradio}.
GNU Radio includes a channel model which simulates atmospheric radio transmission by adding noise,
frequency and timing distortion,
and multi-path echoes caused by radio waves reflecting off various surfaces.
GNU Radio includes generic modulation and demodulation modules which can use an arbitrary constellation,
making this sort of experimentation possible without developing custom modules.

For each constellation type, a plot is shown comparing the Packet Error Rate (PER),
i.e. the ratio of packets that failed to transmit without error, to the Signal to Noise Ratio (SNR), at a fixed blatancy $\beta$.
To show how the blatancy impacts performance,
an additional plot is shown comparing the PER to the blatancy at a fixed SNR,
roughly the signal quality where the secret channel begins to work reliably.
The values shown are the average of 4 separate experiments,
each of which sends 32 128-byte messages.
To avoid obscuring the performance impact of the steganographic channel,
no error correction is used for packet payloads:
even a single bit corrupted counts as a failed packet.
Higher reliability would be simple to achieve in practice,
but the various protocol-specific methods of doing so are outside the scope of this work.

\subsection{Square 4x4-QAM}

\begin{figure}[t]
  \begin{center}
    \includegraphics[width=0.9\linewidth]{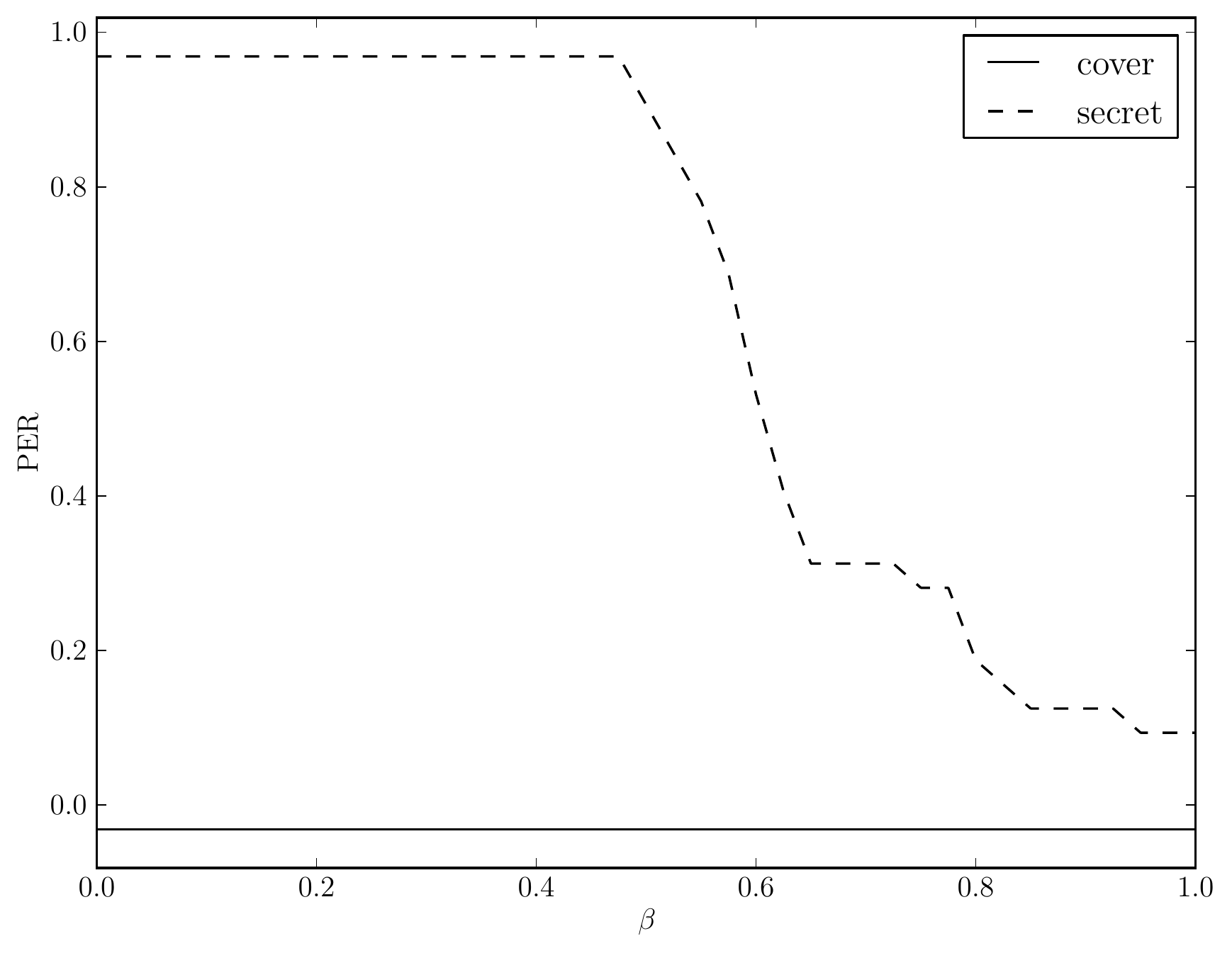}
  \end{center}
  \caption{Square 4x4-QAM Error Rate vs. Blatancy ($SNR = 25 \textnormal{ dB}$)}
  \label{sqam_error_v_blatancy}
\end{figure}

\begin{figure}[t]
  \begin{center}
    \includegraphics[width=0.9\linewidth]{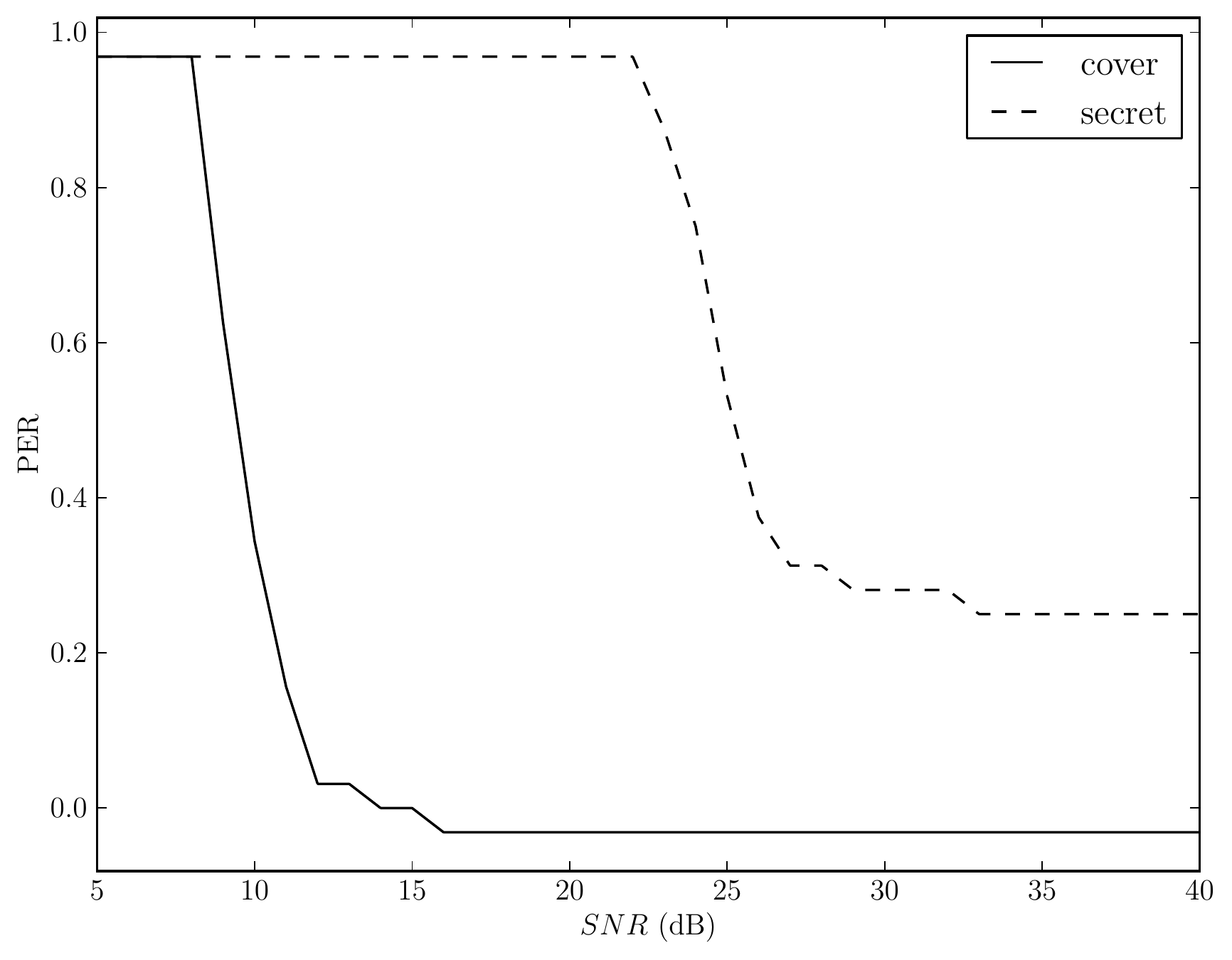}
  \end{center}
  \caption{Square 4x4-QAM Error Rate vs. Signal Quality ($\beta = 0.6$)}
  \label{sqam_error_v_quality}
\end{figure}

The error rate vs. blatancy for square 4x4-QAM is shown in Fig.~\ref{sqam_error_v_blatancy}.
With high blatancy, near perfect transmission of both the cover and secret is achieved.
This is not surprising, since when $\beta = 1.0$ the constellation is the same as standard 16-QAM,
an ideal constellation for 16 symbols.
The signal quality shown here is good enough that the cover transmission works perfectly,
so the cover PER is zero at all blatancies.

The performance over a wide range of signal qualities with $\beta = 0.6$ is shown in Fig.~\ref{sqam_error_v_quality}.
It is clear that the secret channel requires a higher signal quality to function than the cover channel.
This reflects the basic principle of radio steganography: giving up some error tolerance in exchange for a secret channel.

The range where cover transmission is possible but secret transmission is not,
roughly 12 to 20 dB, is clearly visible in Fig.~\ref{sqam_error_v_quality}.
The secret channel is less robust to noise,
but this gap is actually an advantage for steganography.
An observer in this range can reliably receive the cover,
but not the secret, even if the secret constellation is known.
The physical properties of radio waves can provide a level of security beyond what is possible within a reliable medium.
Digital steganography which, for example, embeds hidden information in images,
uses a reliable transport in the hope that an observer does not notice,
but the information is there for any observer to decode.
With radio, secret messages can be exchanged that are impossible to recover for an observer with poor reception,
and to such an observer it appears as if a normal legacy transmission is taking place.
This property suggests many interesting applications,
such as personal area networks with secret channels invisible to an observer,
not via mathematical means but because a distant observer simply does not have sufficient signal
quality to receive the secret channel.

\subsection{Circular 4x4-QAM}

\begin{figure}[t]
  \begin{center}
    \includegraphics[width=0.9\linewidth]{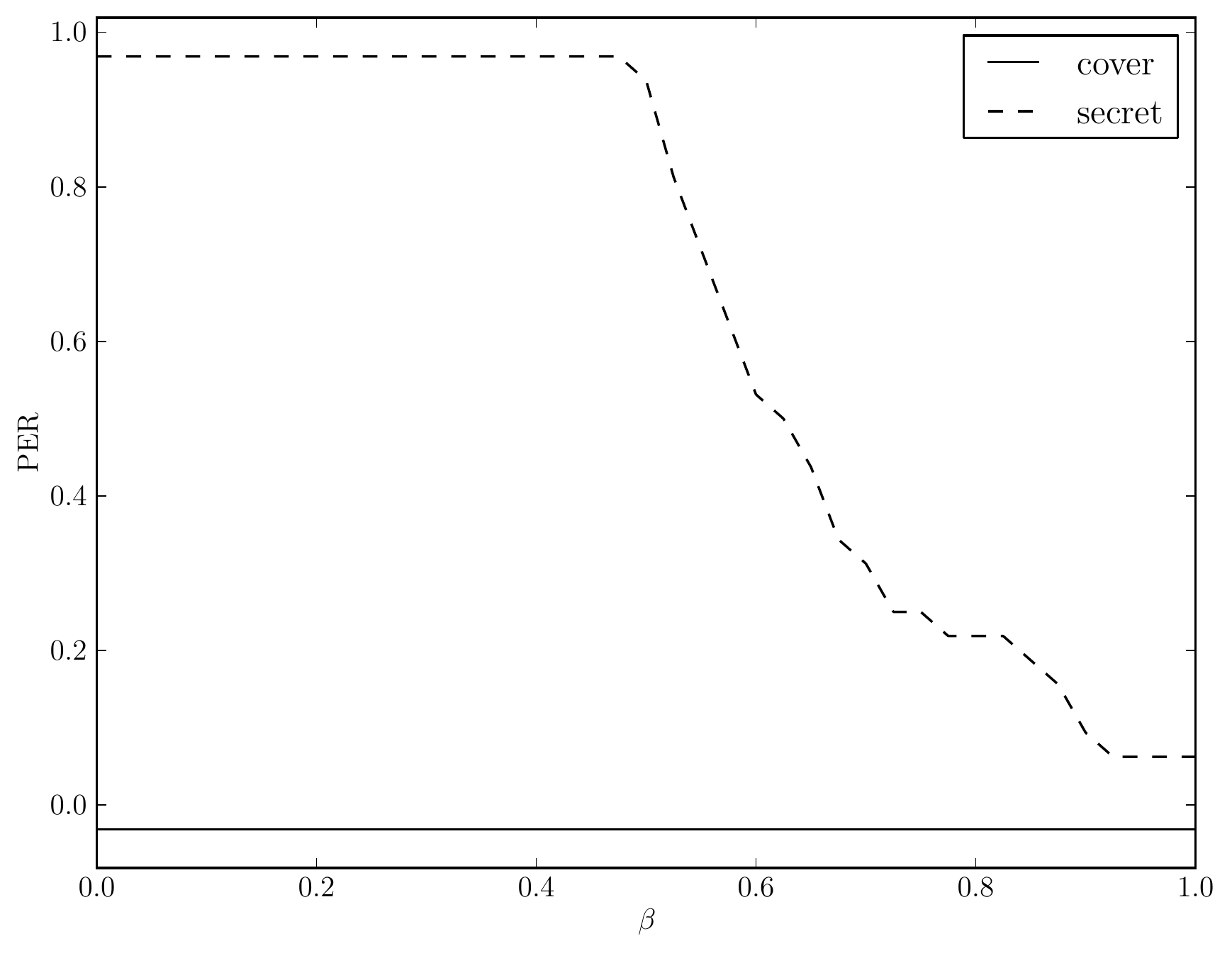}
  \end{center}
  \caption{Circular 4x4-QAM Error Rate vs. Blatancy ($SNR = 25\textnormal{ dB}$)}
  \label{cqam_error_v_blatancy}
\end{figure}

\begin{figure}[t]
  \begin{center}
    \includegraphics[width=0.9\linewidth]{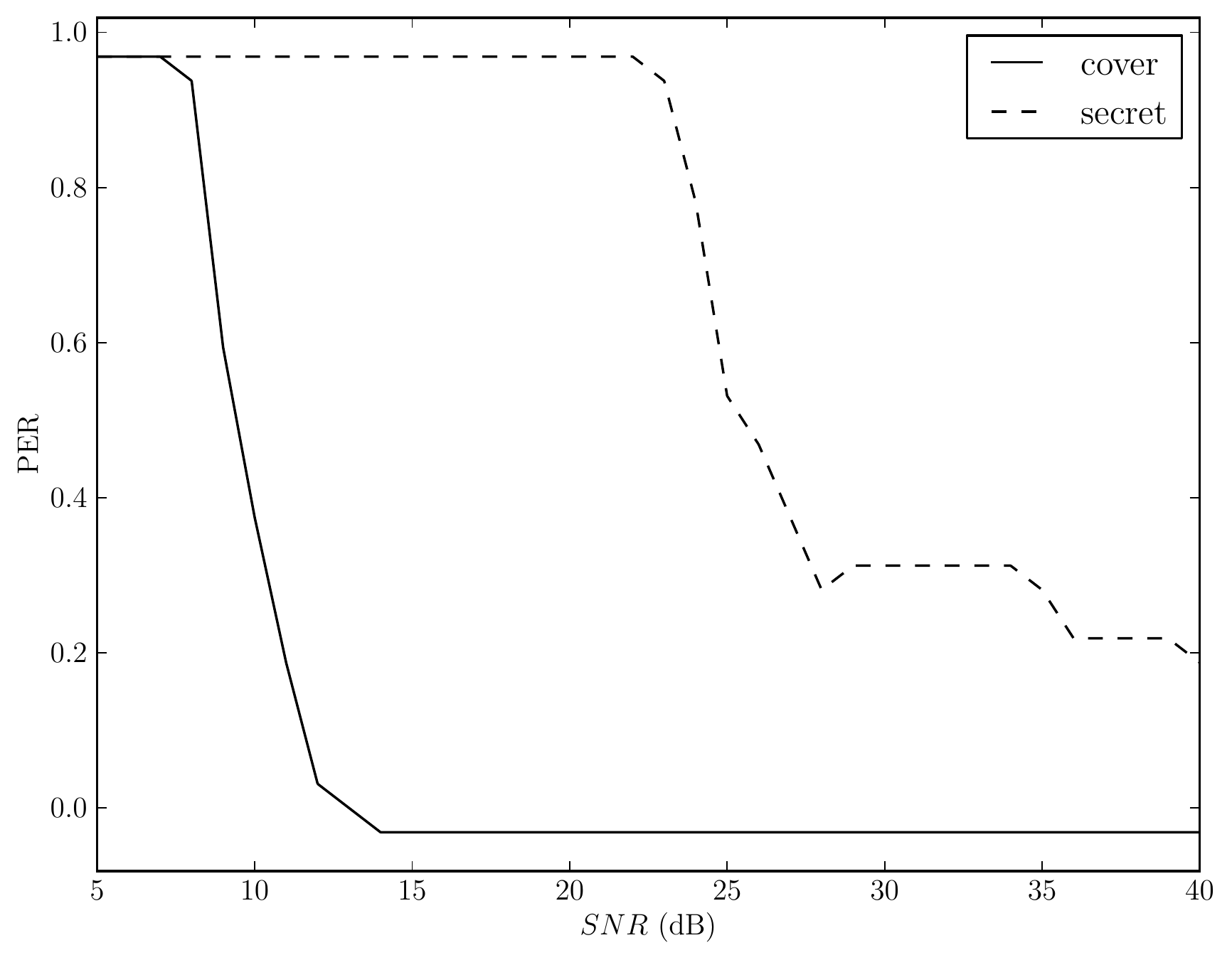}
  \end{center}
  \caption{Circular 4x4-QAM Error Rate vs. Signal Quality ($\beta = 0.6$)}
  \label{cqam_error_v_quality}
\end{figure}

The circular 4x4-QAM constellation is very similar to the 4x4-QAM constellation,
except that each secret symbol cluster is arranged slightly differently.
Accordingly, the performance differences should be minimal.
The PER vs. blatancy is shown in Fig.~\ref{cqam_error_v_blatancy},
and the PER vs. SNR in Fig.~\ref{cqam_error_v_quality}.
Both plots use the same fixed values as the corresponding plots for square 4x4-QAM.

As expected, both arrangements have very similar performance,
but the circular arrangement performs slightly worse with the same parameters.
Roughly, it requires about 0.05 higher blatancy at 25 dB,
or 2 dB better signal quality at $\beta = 0.6$,
to achieve similar performance to square 4x4 QAM.
The error rate also flattens out more slowly as the blatancy is increased,
though at 25 dB this effect is minimal.
The circular variant is slightly less robust since,
unlike the square variant,
$\beta = 1.0$ is somewhat strange and does not correspond to a standard 16 point constellation.
This is because the points are not evenly distributed in order to maximise the distance between any two points,
which is the general ideal for constellation arrangements.

It could be considered a minor advantage that the constellation at high blatancy does not appear to be any standard constellation,
since an observer may attempt to decode an unknown signal by trying every standard constellation.
Since this arrangement does not correspond to any standard constellation,
such a search will not be successful.
This prevents a very simple automatic process from determining the secret constellation,
though is no defense against more sophisticated statistical analysis.

Despite the slightly inferior performance,
circular 4x4-QAM is more convenient than the square arrangement for adding additional modulation in order to increase stealth.
Section~\ref{stealth} discusses the stealth of each constellation type,
and such a modulation of circular 4x4-QAM in detail.
  
\subsection{4x4-PSK}

\begin{figure}[t]
  \begin{center}
    \includegraphics[width=0.9\linewidth]{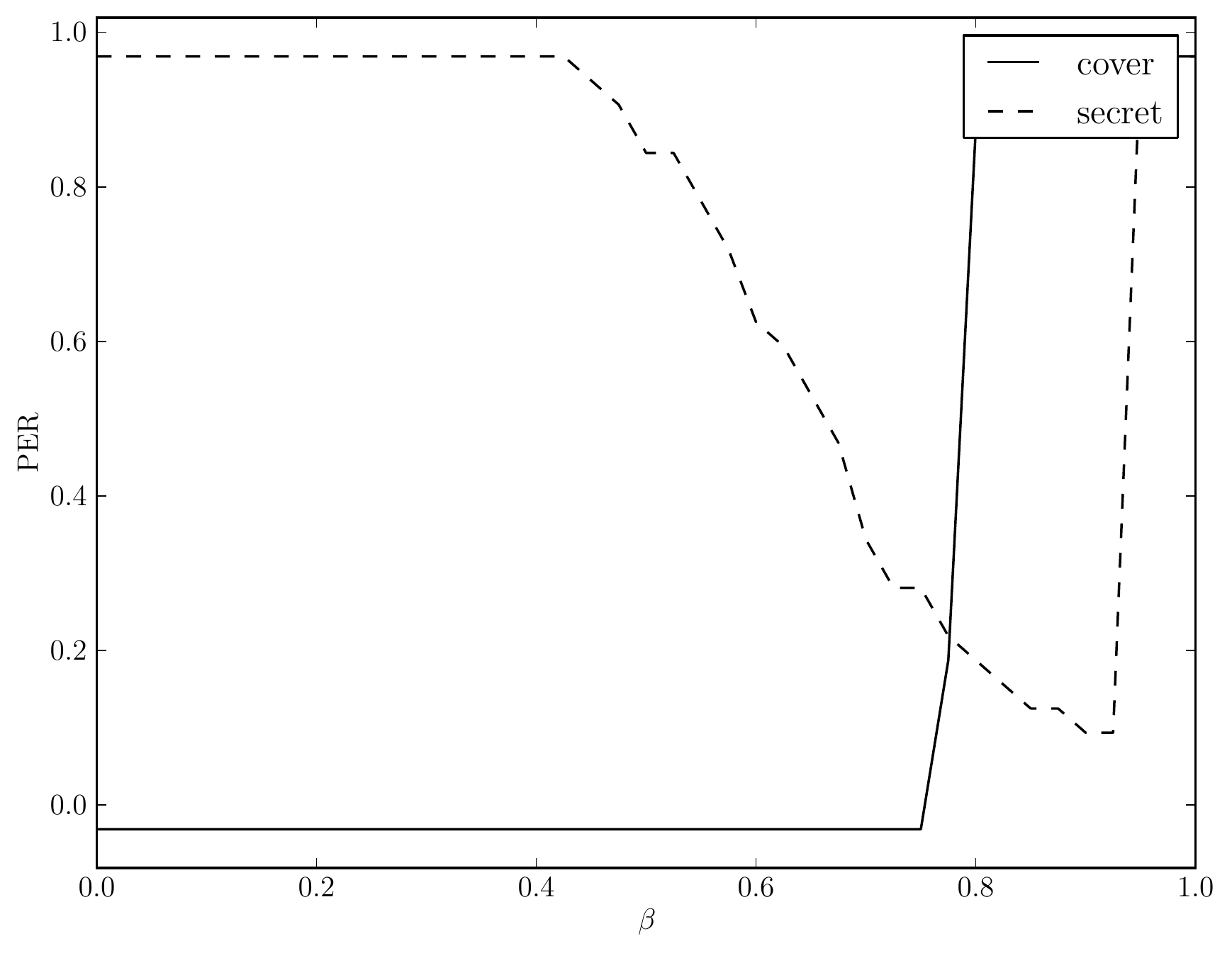}
  \end{center}
  \caption{4x4-PSK Error Rate vs. Blatancy ($SNR = 25 \textnormal{ dB}$)}
  \label{psk_error_v_blatancy}
\end{figure}

\begin{figure}[t]
  \begin{center}
    \includegraphics[width=0.9\linewidth]{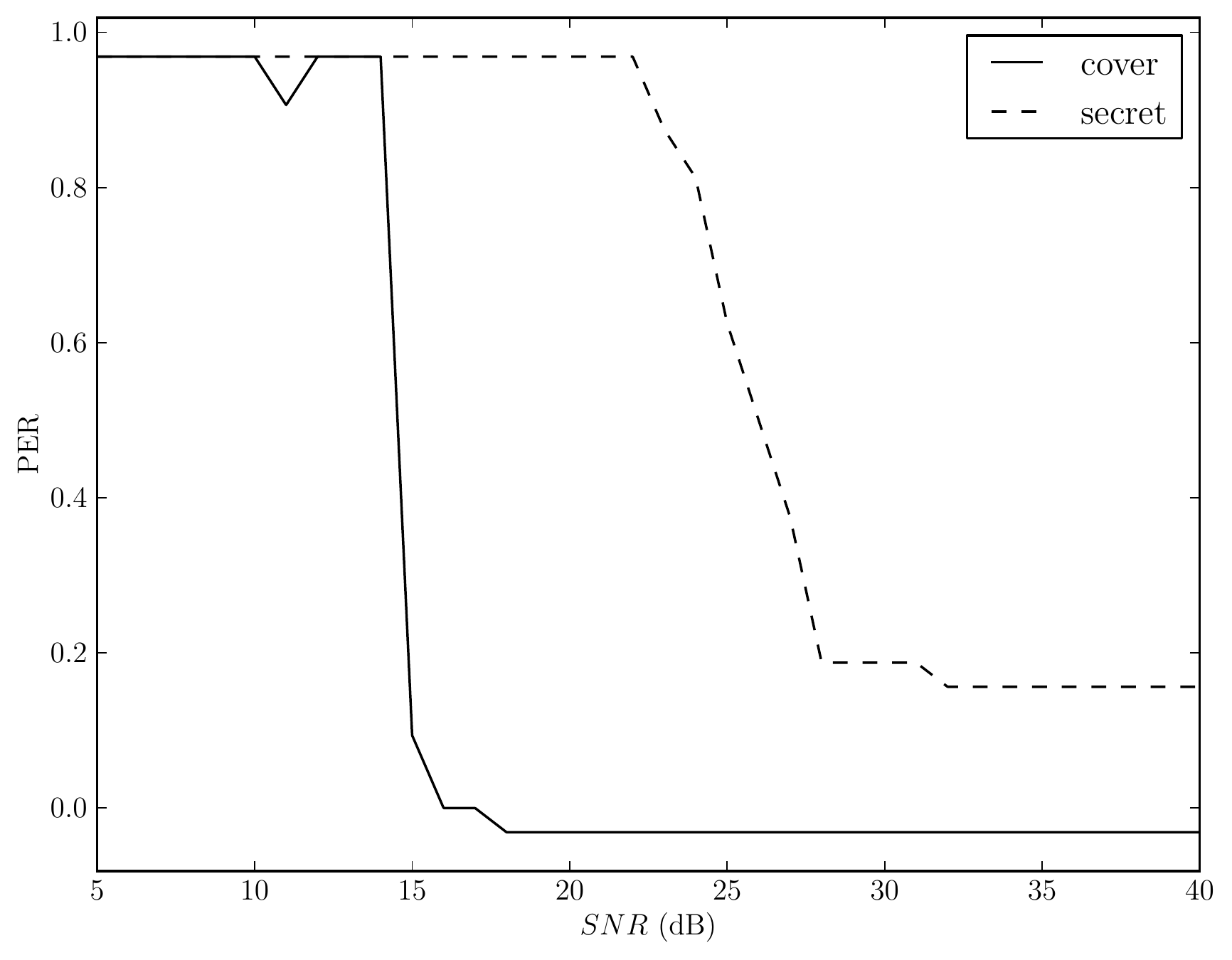}
  \end{center}
  \caption{4x4-PSK Error Rate vs. Signal Quality ($\beta = 0.6$)}
  \label{psk_error_v_quality}
\end{figure}

When only phase modulation is used,
it is not possible to cluster 4 points with equal distance to the original point.
As a result this constellation is not expected to perform as well as the QAM schemes.
The results shown in Fig.~\ref{psk_error_v_blatancy} and Fig.~\ref{psk_error_v_quality} confirm this to be the case.

While $\beta = 1.0$ corresponds to a standard 16-PSK constellation,
some secret symbols are spread very far from their corresponding legacy symbols.
Hence, very high blatancy interferes with the cover much more than in either of the QAM schemes.
Fig.~\ref{psk_error_v_blatancy} shows that this is the case even for the relatively high signal quality of 25 dB.
Here there is a relatively small range of feasible blatancy,
roughly 0.5 to 0.7.
Because of this, tuning would be more difficult with this level of signal quality,
especially in varying conditions,
though an automatic solution would be able to provide a reliable channel.
The same quality, 25 dB, is shown here to facilitate direct comparison with the other constellations,
but realistically a higher SNR would be required for good performance with 4x4-PSK.
As can be seen in Fig.~\ref{psk_error_v_quality},
at around 28 dB near-perfect performance is possible with $\beta = 0.6$.

While the effects of blatancy manipulation here are considerably more sporadic and unpredictable,
the signal does show a predicable improvement in error rate as quality increases.
With higher signal quality the range of feasible blatancies would be wider,
making manual or automatic tuning simpler.

The 4x4-PSK constellation results in a signal that is both more obvious and less robust than either QAM arrangement.
However, these results show that constellation-based steganography is possible in a context where only phase modulation is possible.

\section{Stealth}
\label{stealth}

Due to noise, symbol points as decoded by a receiver differ from the ideal constellation.
The stealth of each scheme can thus be visualised as a scatter plot of the received points.
Such a plot for a stealthy signal looks like a noisy legacy constellation,
but the secret constellation is visible for a less stealthy signal.
Because a transmission is composed of many thousands of symbols,
a histogram is shown on each axis to better illustrate their distribution.
This plot is referred to as the \emph{appearance} of the signal.
The plots shown here are generated from the same signal data used to show performance in the previous section.

\subsection{Square 4x4-QAM}

\begin{figure}[t]
  \begin{center}
    \subfigure[$\beta=0.2$]{
      \includegraphics[width=0.46\linewidth]{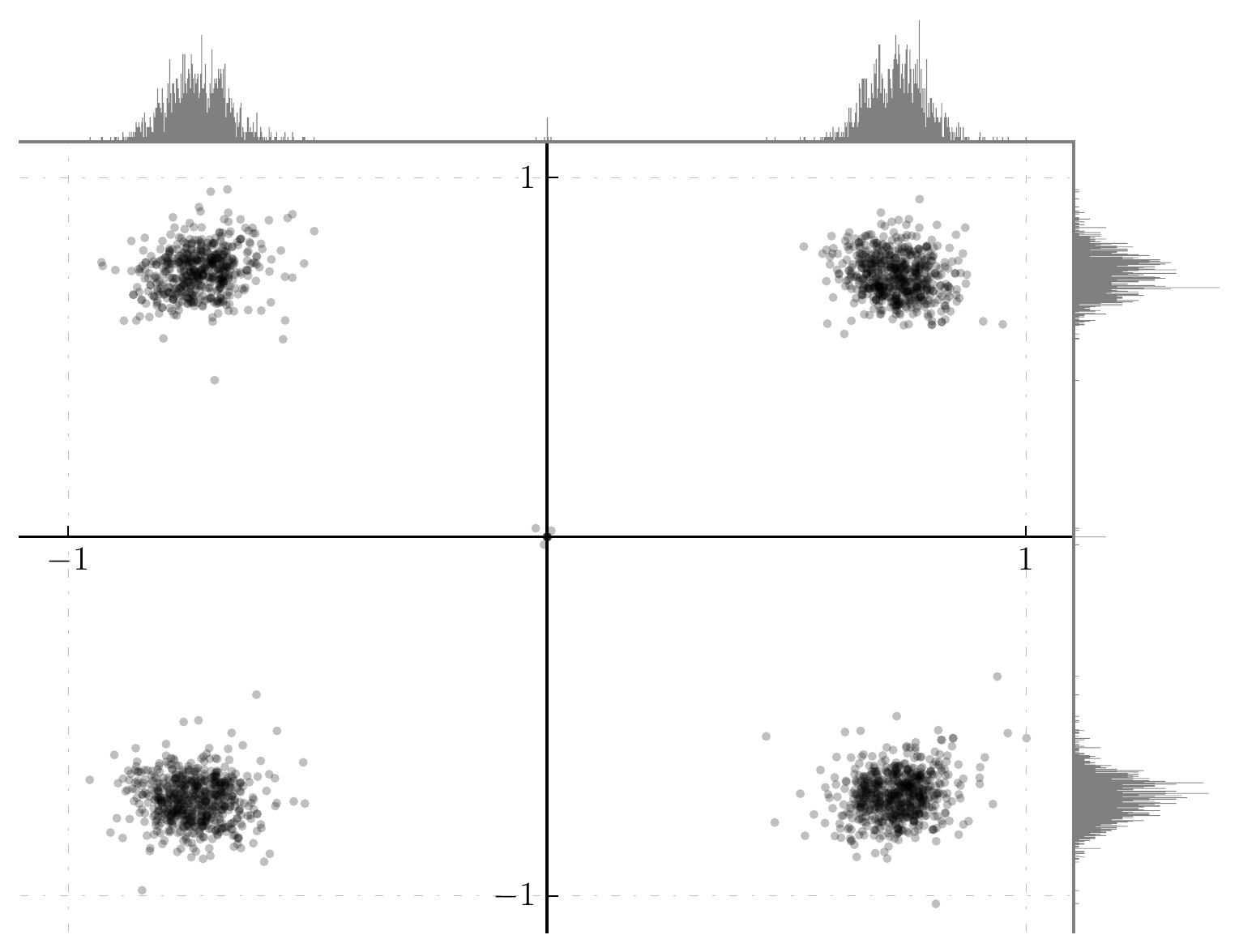}
    }
    \subfigure[$\beta=0.6$]{
      \includegraphics[width=0.46\linewidth]{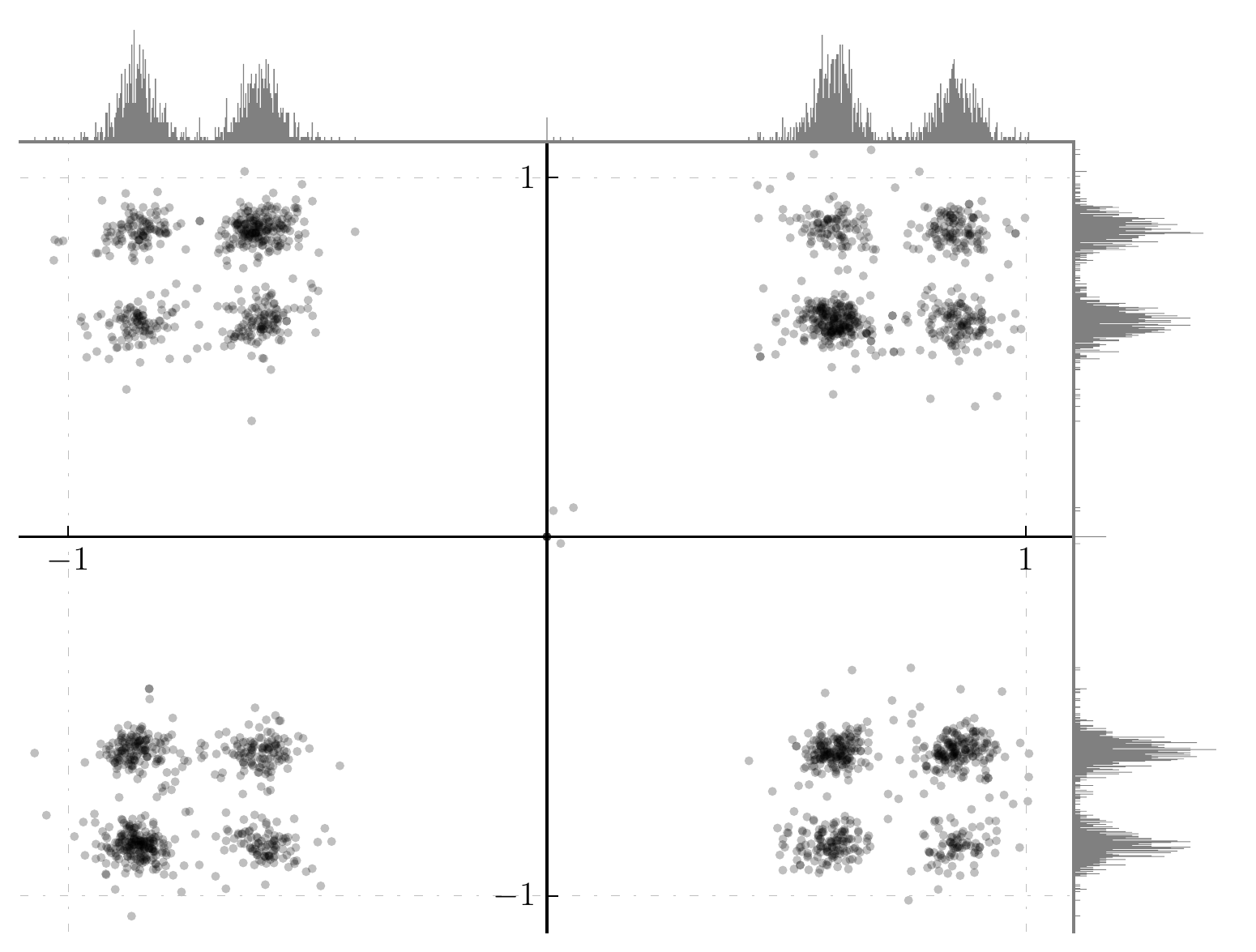}
    }
  \end{center}
  \caption{Square 4x4-QAM Appearance (SNR = 25 dB)}
  \label{sqam_appearance}
\end{figure}

The appearance of the square 4x4-QAM signal is shown in Fig.~\ref{sqam_appearance}.
The histogram at the top and right of each plot shows the frequency of points received at a given position.
At the relatively low $\beta = 0.2$ the transmission is subtle.
The signal appears to be a normal 4-QAM transmission with some noise,
so an observer is unlikely to suspect covert communication.
However, at higher blatancy, like $\beta = 0.6$,
the fact that there are actually 16 points in use is very obvious.
The technique works, but there is much room for improvement on stealth.

\subsection{Circular 4x4-QAM}

\begin{figure}[t]
  \begin{center}
    \subfigure[$\beta=0.2$]{
      \includegraphics[width=0.46\linewidth]{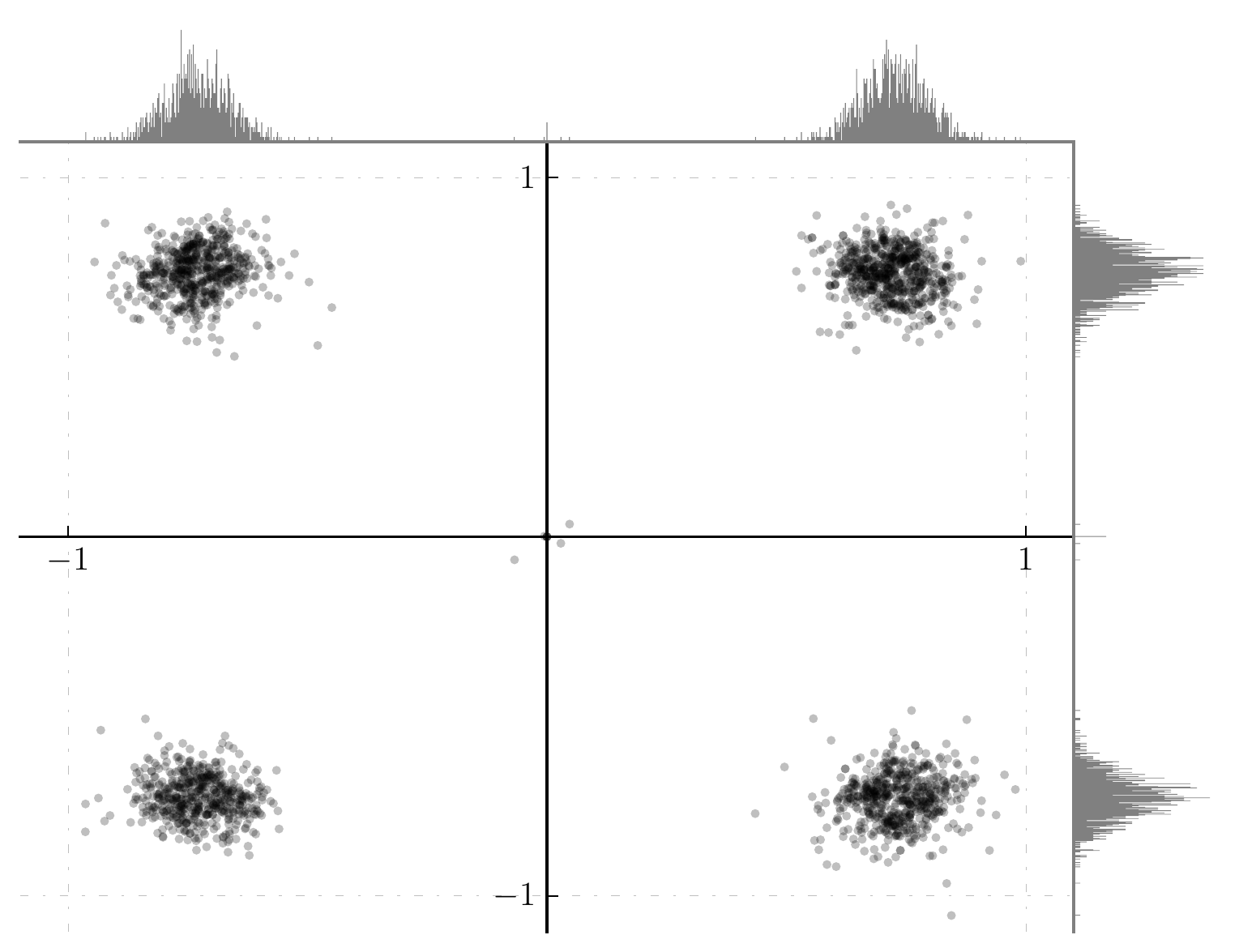}
    }
    \subfigure[$\beta=0.6$]{
      \includegraphics[width=0.46\linewidth]{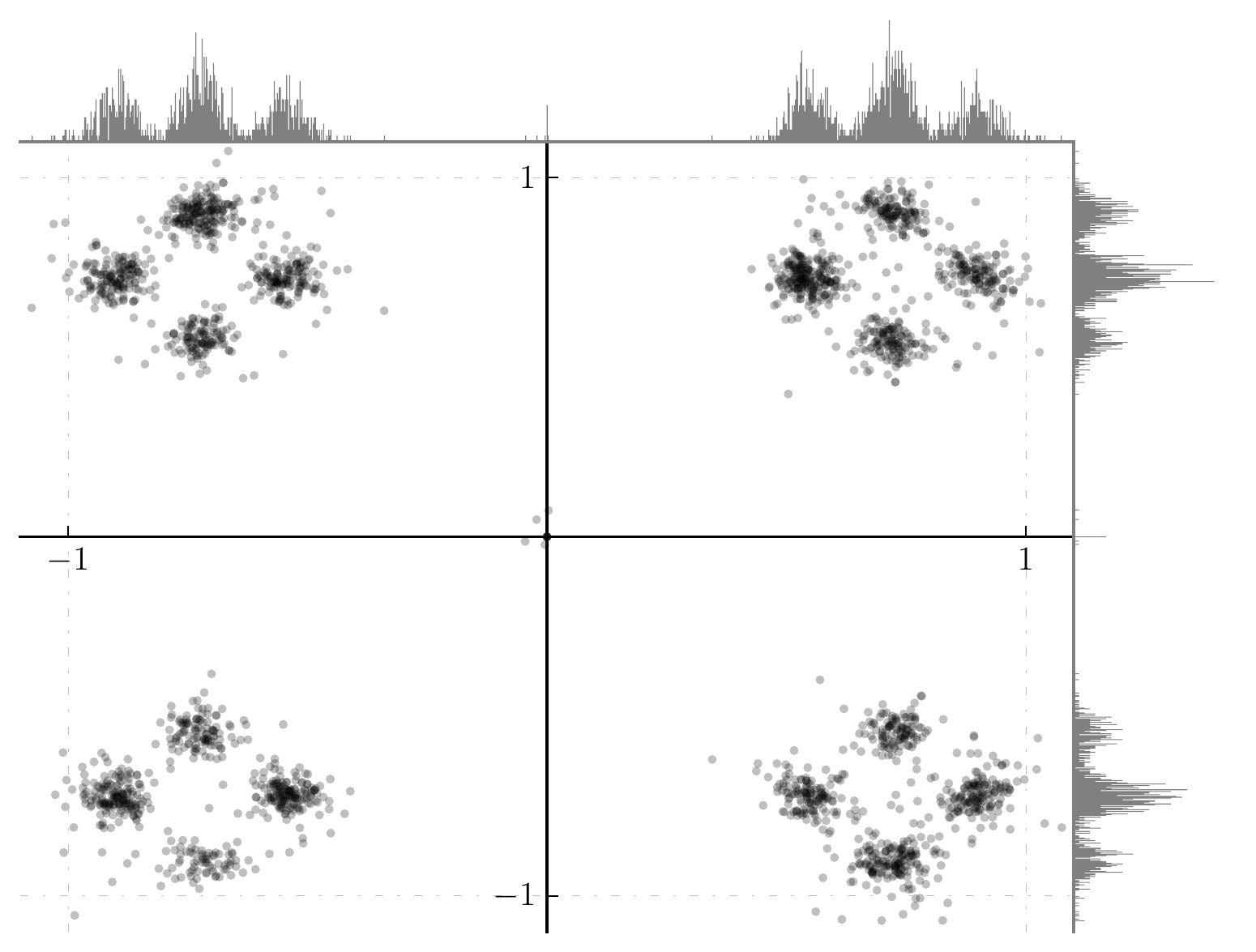}
    }
  \end{center}
  \caption{Circular 4x4-QAM Appearance (SNR = 25 dB)}
  \label{cqam_appearance}
\end{figure}

Comparing the appearance of circular 4x4-QAM shown in Fig.~\ref{cqam_appearance} to the square arrangement is slightly more interesting.
The rotated cluster configuration causes 6 distinct peaks along each histogram axis,
rather than the 4 peaks of the square variant.
Thus it is slightly better than square 4x4-QAM in terms of stealth,
but still quite obvious at this signal quality.
An improvement based on this observation is described in Section~\ref{dynamic}.
As before, at $\beta = 0.2$ the signal appears to be a legacy signal,
though communication with such low blatancy would require higher signal quality.

\subsection{4x4-PSK}

\begin{figure}[t]
  \begin{center}
    \subfigure[$\beta=0.2$]{
      \includegraphics[width=0.46\linewidth]{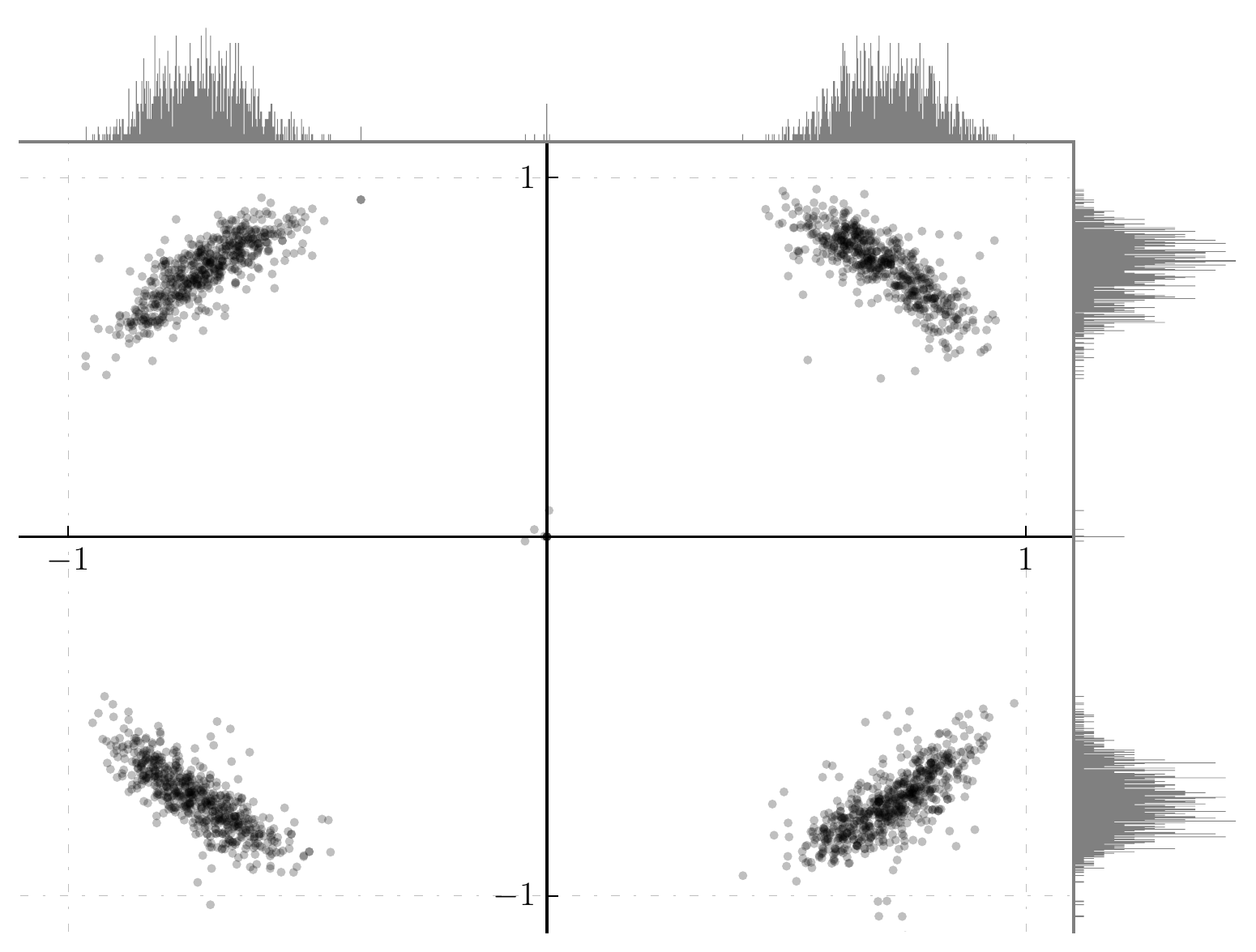}
    }
    \subfigure[$\beta=0.6$]{
      \includegraphics[width=0.46\linewidth]{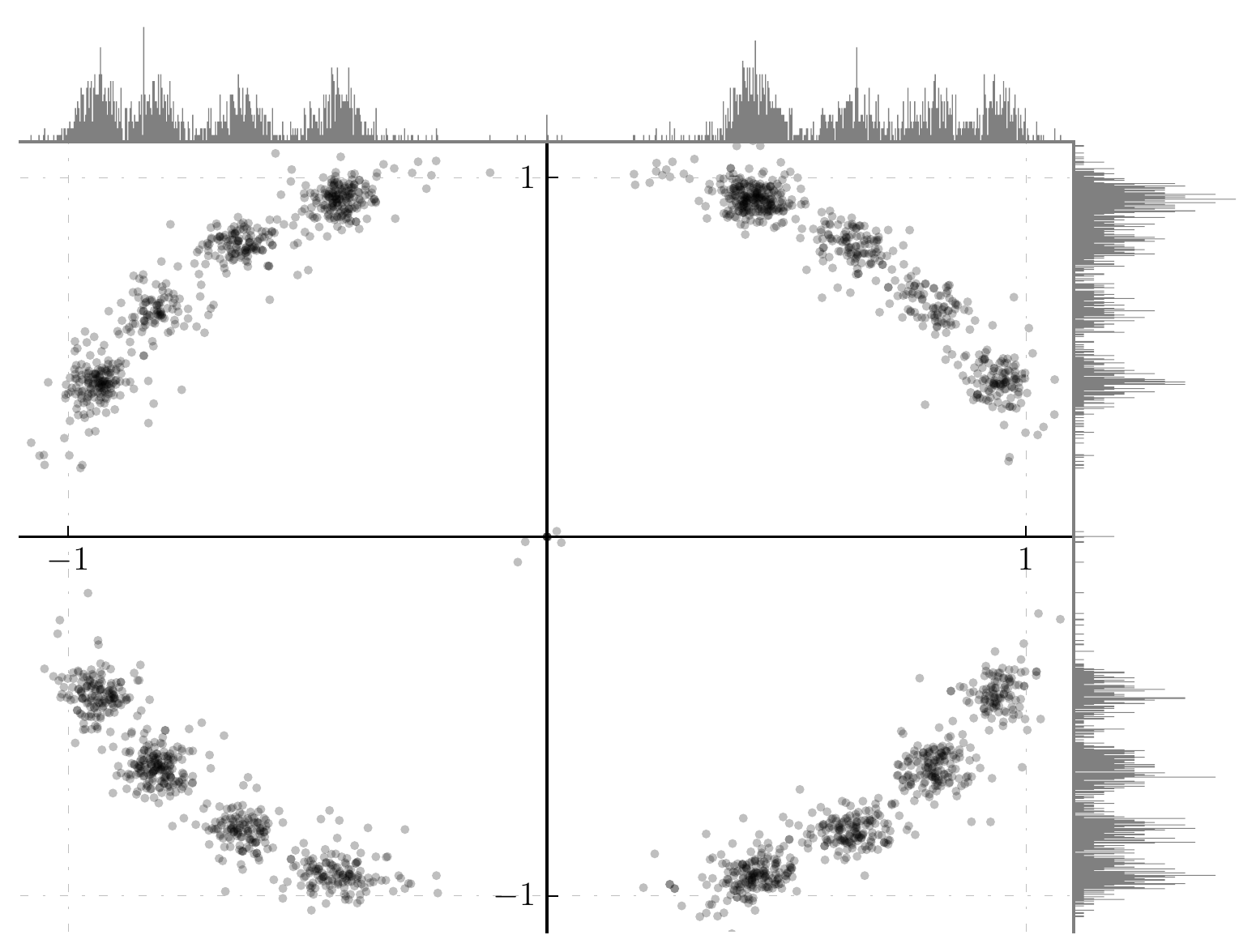}
    }
  \end{center}
  \caption{4x4-PSK Appearance (SNR = 25 dB)}
  \label{psk_appearance}
\end{figure}

The appearance shown in Fig.~\ref{psk_appearance} confirms the expectation that the PSK scheme is the least stealthy due to the wide spread of secret symbols.
At $\beta = 0.6$ the secret constellation is obvious as with the other schemes,
but here even at $\beta = 0.2$ it is clear that something other than normal 4-PSK
is being used.
Atmospheric noise tends to scatter points in a circular region around the ideal points.
Here points are spread in a distinct arc, which is unlikely to occur as a result of noise or other environmental conditions.
Thus the signal is more obvious than either QAM arrangement,
since symbols are not distributed evenly about their corresponding legacy symbols.

\subsection{Dynamic Constellations}
\label{dynamic}

The existence of the secret channel is obvious to an observer with sufficiently high $SNR$
since the secret symbols are visibly distinct when the signal is analysed over time.

This situation can be improved by dynamically modulating the constellation such that the positions of secret symbols vary over time,
but remain clustered around the original point.
The circular 4x4-QAM scheme is best suited to this purpose;
each secret point is arranged in a circle around the cover point,
so the angle at which they are placed can be modulated to achieve the desired effect.

This dynamic modulation is defined in terms of a new parameter, the \emph{shift} $\sigma$.
The shift is normalised to $[0,1]$ such that 0 is the original secret constellation,
which is adjusted as the shift increases in a circular fashion such that 1 is also the original secret constellation.
This prevents sudden dramatic changes in the constellation,
which has a negative impact on performance.
As transmission occurs, the shift at step $t$ is $\sigma_t = \sigma_{t - 1} + \epsilon \pmod{1}$
for some $0 < \epsilon \ll 1$.
For circular 4x4-QAM, the shift corresponds to an angular offset where $\sigma = 0 \Rightarrow 0$ radians,
and $\sigma = 1 \Rightarrow 2\pi$ radians.
This results in each cluster making a complete revolution as $\sigma$ increasees from 0 to 1.

The blatancy is similarly modulated within a range in order to spread the points radially as well as angularly.
The $\beta$ parameter then becomes a range; the blatancy oscillates within this range during transmission.

Unlike the static secret constellations,
this method requires some form of synchronisation between the transmitter and receiver.
If the constellations become unsynchronised,
secret communication becomes impossible.
Any synchronisation method is sufficient,
for example a slow modulation based on a shared real-time clock (e.g. GPS or NTP),
or increasing the shift by a known amount when the sender receives a successful acknowledgement of packet reception.

The GNU radio modulation and demodulation modules were not designed for dynamic constellation manipulation.
In order to evaluate the performance of this method,
the C++ implementations of the modulation and demodulation modules were modified to allow changing the constellation without destroying the module.
Replacing the module instances every time the constellation changes does not work,
since this destroys the internal time synchronisation state\footnote{
Note this is unrelated to the high level synchronisation requirements of dynamic constellations.}
required for symbol decoding.

\begin{figure}[t]
  \begin{center}
    \includegraphics[width=0.9\linewidth]{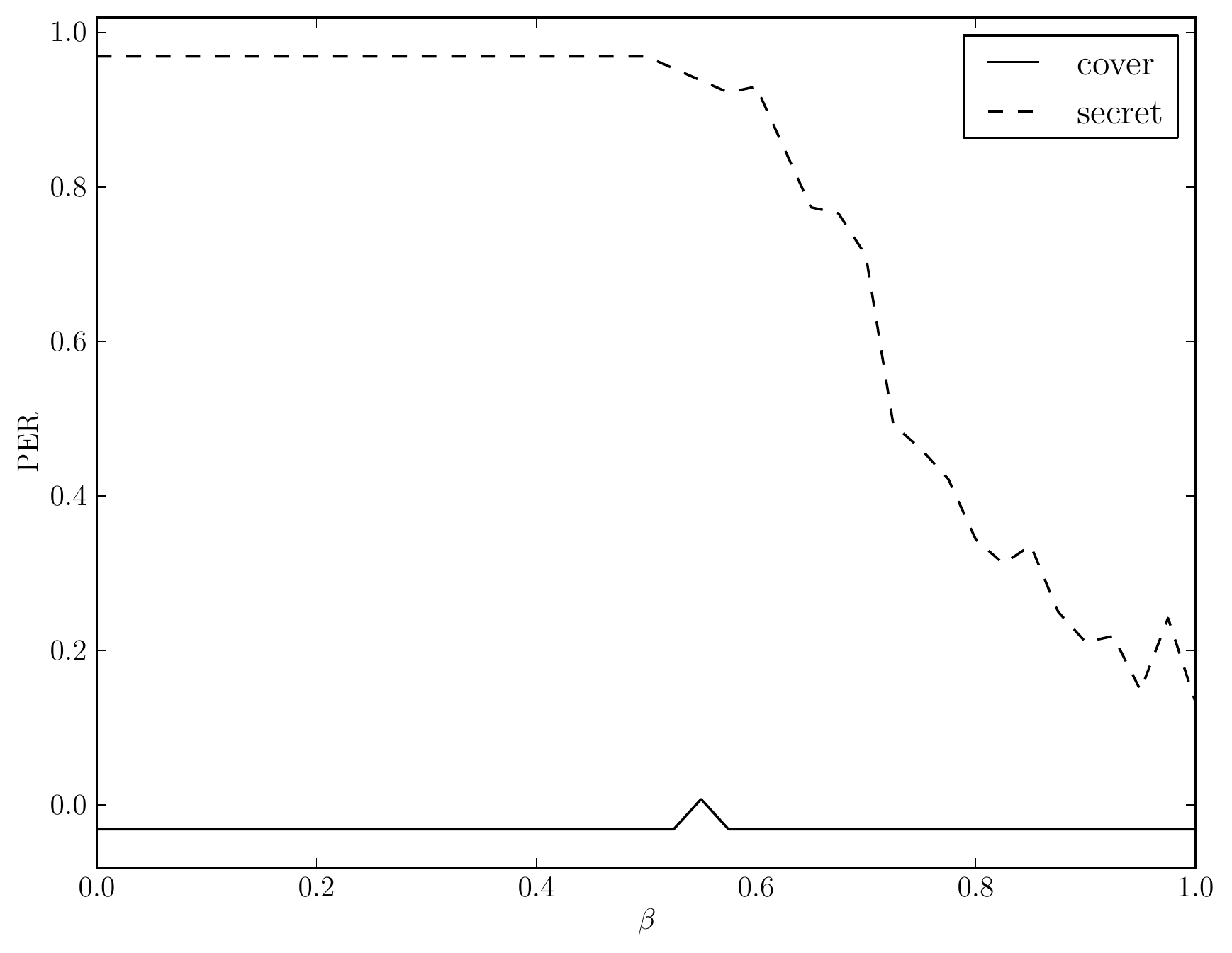}
  \end{center}
  \caption{Shifted 4x4-QAM Error Rate vs. Blatancy (SNR = 25 dB)}
  \label{vqam_error_v_blatancy}
\end{figure}

\begin{figure}[t]
  \begin{center}
    \includegraphics[width=0.9\linewidth]{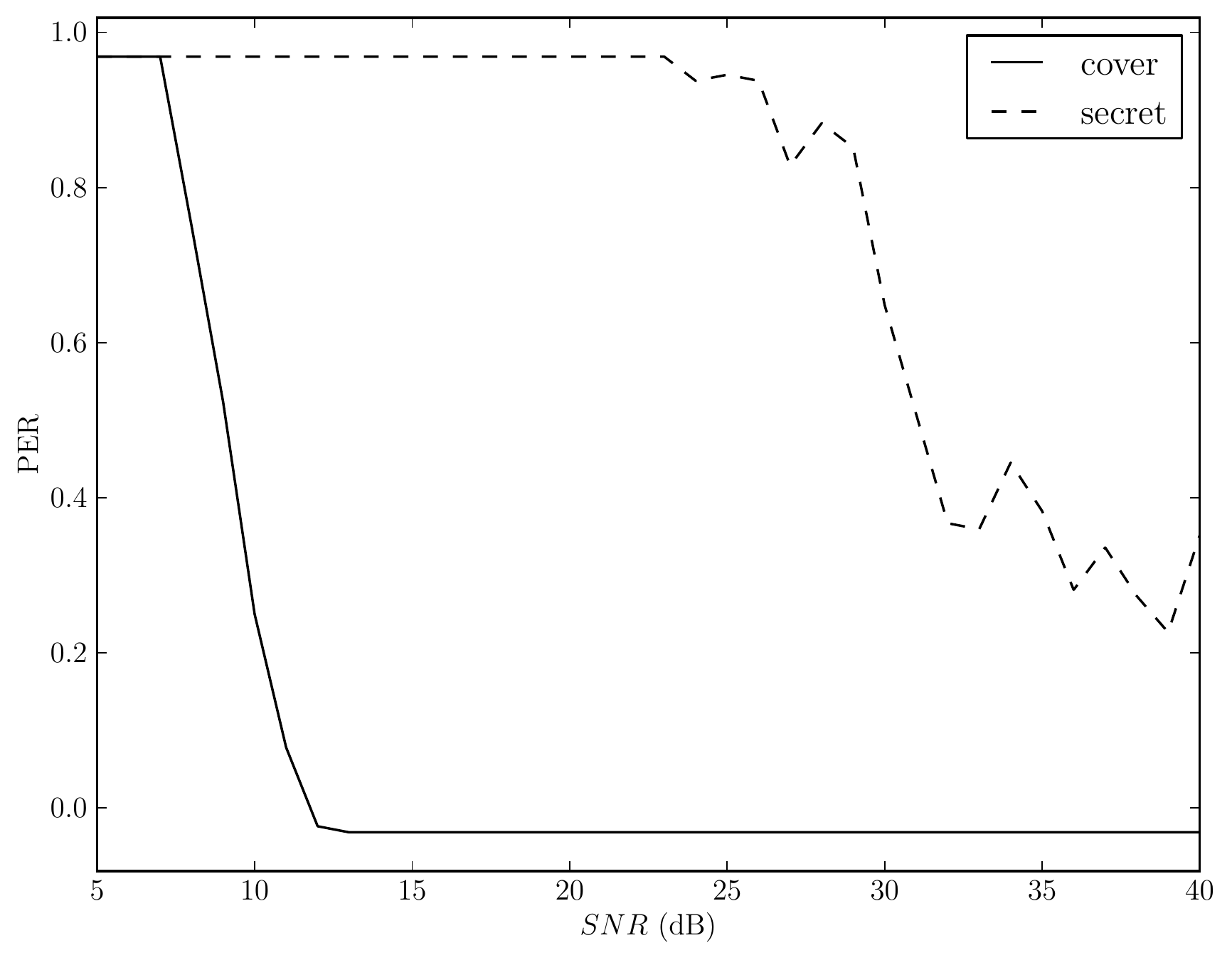}
  \end{center}
  \caption{Shifted 4x4-QAM Error Rate vs. Signal Quality ($\beta = 0.6$)}
  \label{vqam_error_v_quality}
\end{figure}

\begin{figure}
  \subfigure[Shifted, SNR = 25 dB]{
	\resizebox{0.45\linewidth}{!}{
      \includegraphics{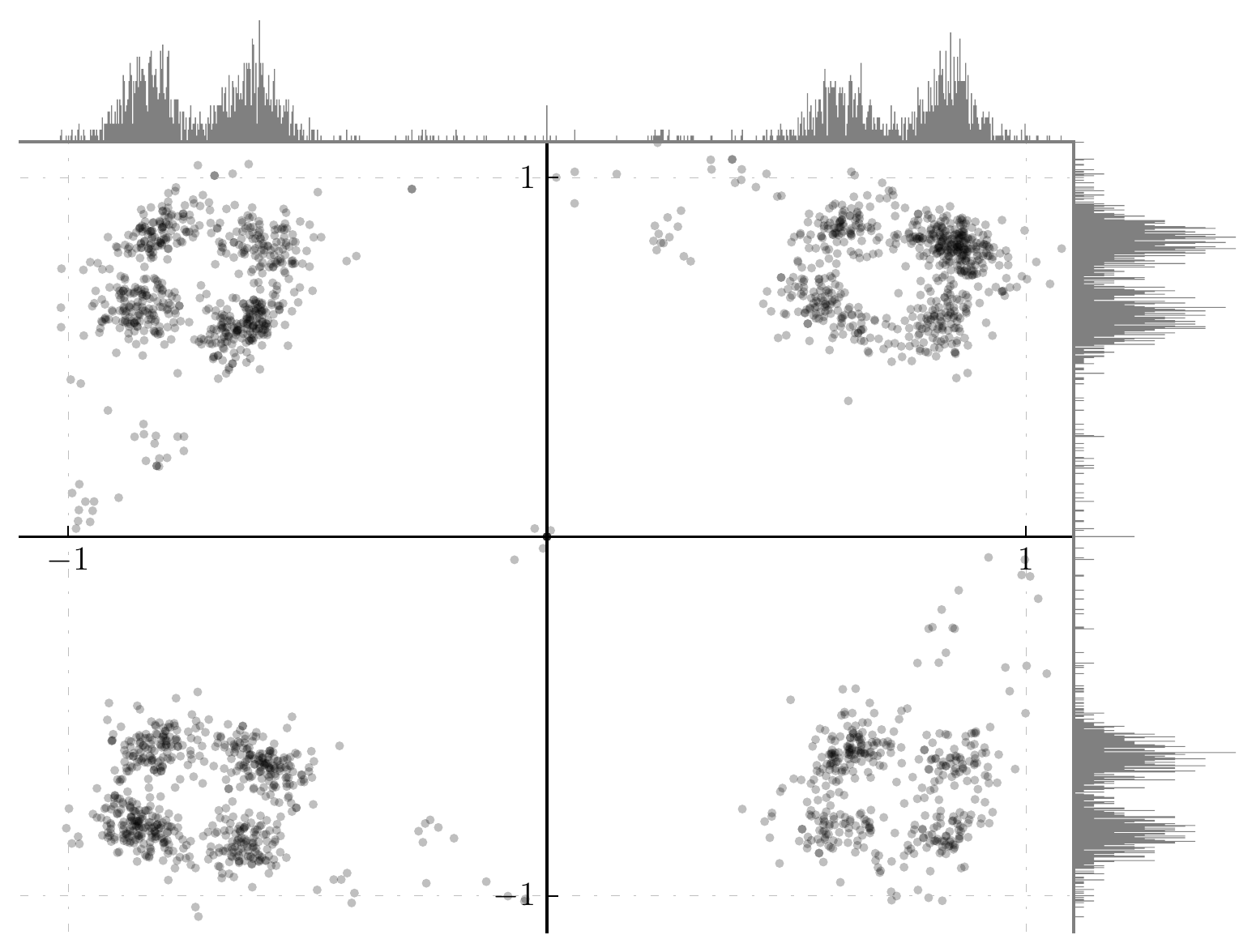}
    }
  }
  \subfigure[Shifted, SNR = 20 dB]{
	\resizebox{0.45\linewidth}{!}{
      \includegraphics{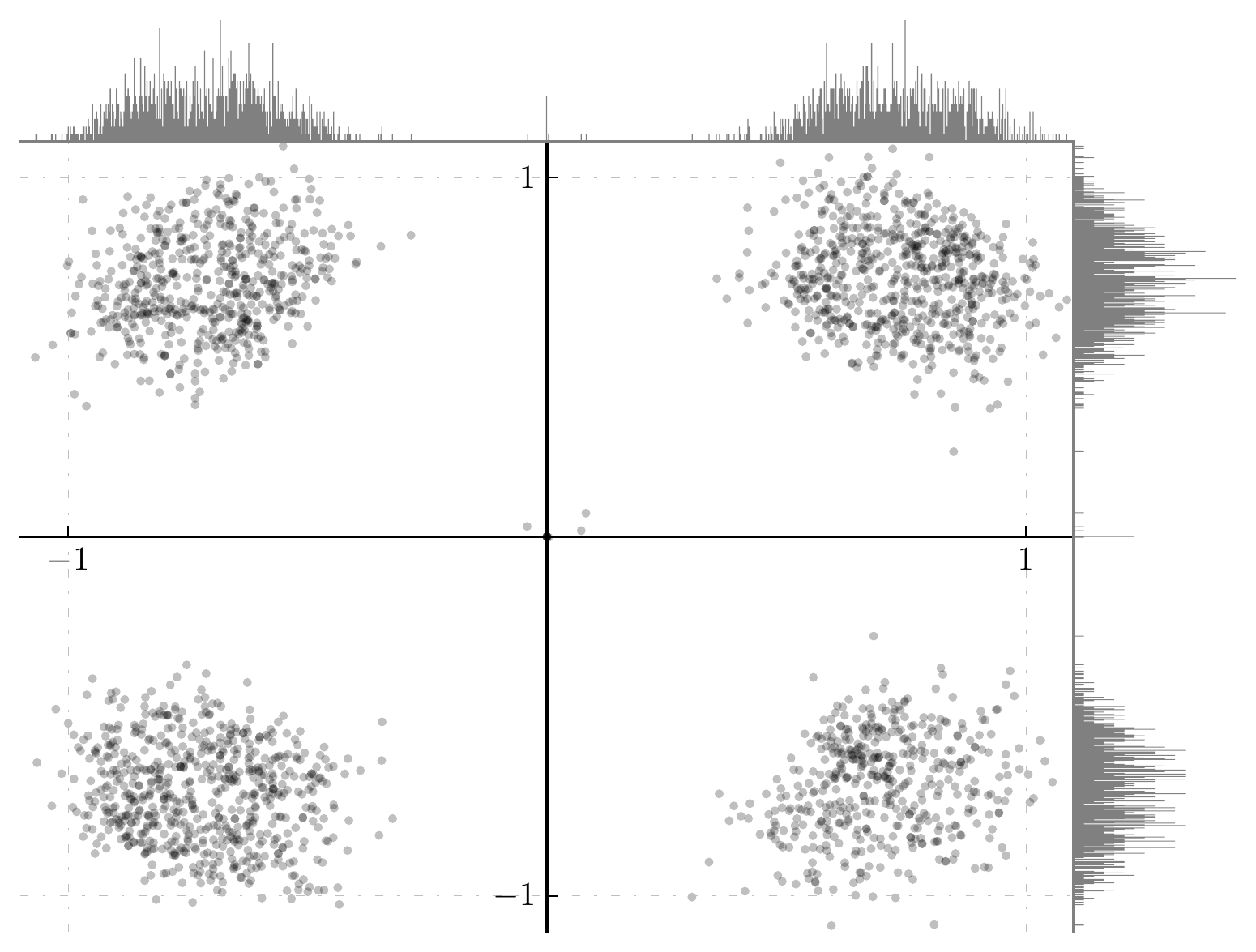}
    }
  }
  \caption{Shifted 4x4-QAM Appearance, $\beta=0.5 \dots 0.7$}
  \label{vqam_appearance}
\end{figure}

The performance of dynamically modulated circular 4x4-QAM is shown in Fig.~\ref{vqam_error_v_blatancy} and Fig.~\ref{vqam_error_v_quality}.
The impact of adjusting the constellation during transmission is clear.
The most distinct effect is more sporadic performance;
the error rate curves are less smooth though the variance of this ``choppiness'' is low enough to be of little concern.

The increased blatancy and signal quality requirements are more significant.
At 25 dB a blatancy about 0.1 higher is required for reliable transmission.
With a blatancy of 0.6, a SNR of about 33 dB is required for reliable transmission,
in contrast to only about 27 dB for the static arrangement,
a difference of 5 dB.
While dynamic modulation clearly has a negative impact on performance,
transmission is still feasible,
and some performance degradation is expected as the cost of increased stealth.
However, it is likely that a modulator and demodulator specifically designed to better tolerate this situation would result in better performance.

The appearance of this signal is shown in Fig.~\ref{vqam_appearance}.
With a high enough SNR to receive the signal,
the points now appear as rings rather than distinct points.
As can be seen in the histograms, this is a considerable improvement in stealth.
While the rings make it clear something other than standard 4-QAM is in use to an observer with sufficiently high SNR,
the secret constellation is not visible since the points change over time.

At the slightly lower SNR of 20 dB,
the rings are much more subtle; the signal looks more like standard 4-QAM with noise.
So, while not very subtle to an observer with a high enough signal quality to receive the secret transmission,
the technique is covert to observers with lower signal quality.

\section{Conclusions and Future Work}

The constellation-based approach to radio steganography presented here is a feasible one with several unique benefits.
Simulations confirm that reliable transmission of both the cover and secret channels is possible,
and constellations can easily be tuned to be as subtle as possible while achieving communication in a given environment.
This method also has the desirable quality,
unique to analogue transmission,
that the secret channel is unrecoverable by an observer with poor signal quality even if the information hiding technique in use is known.

However, the resulting signals are not stealthy enough to evade statistical analysis by an observer with sufficiently high signal quality.
Modulating the constellation over time improves this situation considerably,
at the cost of some performance.
Further work is required to make GNU radio suitable for this type of modulation.
Because of this, and the fact that some form of protocol-dependant synchronisation is required,
it is expected that a real implementation would see better results for this technique.

The methods presented here work well for secret communication with minimal interference to a legacy receiver,
but there are other potential improvements for stealth.
One obvious problem is the ring-like appearance of the dynamically modulated signal.
Possible solutions include a perturbation of circular clusters,
or an alternative ``hubbed'' constellation where the original point is preserved in the secret constellation and surrounded by $k-1$ points.
This would decrease the distance between symbols,
but make the secret channel much more resistant to statistical analysis.

Finally, the approach here is easily applicable to most constellations.
It would be useful to investigate secret constellations for others
such as circular QAM or Frequency Shift Keying (FSK).
Constellations with more than 16 points would also be useful,
and allow for the possibility of increased adaptability by choosing how many points to use in the secret constellation.
For example, it is possible to use a secret 64-QAM constellation with a 4-QAM cover,
though the signal quality requirements for the secret would be much higher.

This preliminary investigation shows that secret constellations are a promising method of hiding information in a digital radio signal,
and possibly in other mediums based on periodic waveforms.
Future work in this area is likely to result in robust and practical covert communication systems.

% Trigger a \newpage before the given reference to balance last page columns
%\IEEEtriggeratref{3}

\bibliographystyle{IEEEtran}
\bibliography{IEEEabrv,Adaptive_Software_Radio_Steganography}

\end{document}